\documentclass[sigconf]{acmart}
\AtBeginDocument{%
  \providecommand\BibTeX{{%
    \normalfont B\kern-0.5em{\scshape i\kern-0.25em b}\kern-0.8em\TeX}}}

\setcopyright{acmcopyright}
\copyrightyear{2018}
\acmYear{2018}
\acmDOI{XXXXXXX.XXXXXXX}

\acmConference[Conference acronym 'XX]{Make sure to enter the correct
  conference title from your rights confirmation emai}{June 03--05,
  2018}{Woodstock, NY}
\acmPrice{15.00}
\acmISBN{978-1-4503-XXXX-X/18/06}

\usepackage{bm}
\usepackage{amsfonts}
\usepackage{amsmath}
\def\method{PolyCF}

\usepackage{multirow}
\usepackage{enumitem}
\usepackage{tabularx}
\newcolumntype{Y}{>{\centering\arraybackslash}X}
\usepackage{graphicx}
\usepackage{subcaption}
\usepackage{booktabs}
\usepackage{algorithm}
\usepackage{algorithmic}
\usepackage[algo2e,ruled,linesnumbered]{algorithm2e}

\begin{document}

\title{PolyCF: Towards the Optimal Spectral Graph Filters for Collaborative Filtering}

\author{Yifang Qin}
\affiliation{%
  \institution{Peking University}
  % \streetaddress{P.O. Box 1212}
  \state{Beijing}
  \country{China}
}
\email{qinyifang@pku.edu.cn}
% \orcid{0000-0002-7520-8039}

\author{Wei Ju}
\affiliation{%
  \institution{Peking University}
  \state{Beijing}
  \country{China}
}
\email{juwei@pku.edu.cn}

\author{Xiao Luo}
\affiliation{%
  \institution{University of California, Los Angeles}
  \state{California}
  \country{USA}
}
\email{xiaoluo@cs.ucla.edu}

\author{Yiyang Gu}
\affiliation{%
  \institution{Peking University}
  \state{Beijing}
  \country{China}
}
\email{yiyanggu@pku.edu.cn}

\author{Zhiping Xiao}
\affiliation{%
  \institution{University of California, Los Angeles}
  \state{California}
  \country{USA}
}
\email{patriciaxiao@g.ucla.edu}

\author{Ming Zhang}
\affiliation{%
  \institution{Peking University}
  \state{Beijing}
  \country{China}
}
\email{mzhang\_cs@pku.edu.cn}

\renewcommand{\shortauthors}{Trovato and Tobin, et al.}


\begin{abstract}
Collaborative Filtering (CF) is a pivotal research area in recommender systems that capitalizes on collaborative similarities between users and items to provide personalized recommendations. With the remarkable achievements of node embedding-based Graph Neural Networks (GNNs), we explore the upper bounds of expressiveness inherent to embedding-based methodologies and tackle the challenges by reframing the CF task as a graph signal processing problem. To this end, we propose \method{}, a flexible graph signal filter that leverages polynomial graph filters to process interaction signals. \method{} exhibits the capability to capture spectral features across multiple eigenspaces through a series of Generalized Gram filters and is able to approximate the optimal polynomial response function for recovering missing interactions. A graph optimization objective and a pair-wise ranking objective are jointly used to optimize the parameters of the convolution kernel. Experiments on three widely adopted datasets demonstrate the superiority of \method{} over current state-of-the-art CF methods. Moreover, comprehensive studies empirically validate each component's efficacy in the proposed \method{}.

\end{abstract}






\maketitle

\section{Introduction}

Collaborative Filtering (CF) stands out as one of the most popular research topics of recommender systems, offering an effective solution to the issue of information overload in a wide range of web applications \cite{covington2016deep,wang2018billion,he2020lightgcn}. The primary goal of CF tasks is to recommend the most suitable items for each user to interact with based on their historical interactions. Early CF methods have traditionally framed the CF task as a matrix factorization problem \cite{mnih2007probabilistic,koren2009matrix,gopalan2015scalable}. Subsequent research has emphasized that the crux of solving CF problems lies in modeling the collaborative affinity between users and items according to their interaction history \cite{he2017neural,ebesu2018collaborative}.

With the rapid advancement of Graph Neural Networks (GNNs) and their notable success in graph representation and other downstream tasks \cite{defferrard2016convolutional,kipf2016semi,hamilton2017inductive,ju2023comprehensive}, an increasing number of research endeavors have shifted their focus toward applying graph-based methods to Collaborative Filtering. Early works such as NGCF \cite{wang2019neural} and LightCGN \cite{he2020lightgcn} integrate graph convolution networks (GCNs) into interaction graphs to capture high-order connectivity between users and items. These graph-based models map the nodes into embedding vectors and fully exploit neighborhood structures through a message-passing paradigm. Inspired by these efforts, later graph-based models have further enhanced the model performance and expressiveness of graph embedding-based CF methods.

For instance, DGCF \cite{wang2020disentangled} introduces disentangled graph convolution to capture rich semantics of interactions. UltraGCN \cite{mao2021ultragcn} extends the propagation function of LightGCN to an infinite number of layers while simplifying the computation process. JGCF \cite{guo2023manipulating} focuses on applying trainable linear polynomial graph convolution to capture spectral features. Other works concentrate on enhancing the learning of more representative node embeddings through contrastive learning methods. SGL \cite{wu2021self} proposes to learn informative node representations through contrastive learning between original and augmented graphs. LightGCL \cite{cai2023lightgcl} further enhances the contrastive view with the singular value decomposition of the interaction graphs, offering a spectral perspective.

Despite the prosperity of node embedding-based graph CF models, there are increasing recent researches revealing the limitations of such methods and achieving the state-of-the-art performance through alternative approaches. EASE\textsuperscript{R} \cite{steck2019embarrassingly} discusses the closed-form solution of the training objective for a linear encoder and achieves unexpectedly promising results with a shallow one-layer filter. Later works focus on obtaining powerful graph filters and tackling CF problems using graph signal processing. GF-CF \cite{shen2021powerful} and PGSP \cite{liu2023personalized} explore the effectiveness of graph low-pass filters as a universal solution for various collaborative filtering tasks. GS-IMC \cite{chen2023graph} introduces a class of regularization functions to address CF as an inductive one-bit matrix completion task. However, it is essential to note that existing graph signal processing methods heavily rely on manually crafted graph filters, often derived from empirical observations made with previous CF models. Furthermore, the theoretical analysis of these utilized filters is currently limited to symmetrical normalized Laplacian matrices, which consequently restricts the potential of the adopted graph filters.

To address the aforementioned challenges of graph-based CF models, this research commences with a comprehensive analysis of the upper bounds of expressiveness achievable by node embedding-based graph models. Subsequently, we introduce the rationale behind a novel graph filter-based approach, denoted as \method{}, which goes beyond limitations imposed by the embedding size of nodes. 
We equip \method{} with a novel generalized normalization of the Gram matrix as the filter backbone, thereby enabling it to capture spectral characteristics originating from diverse eigenspace structures. A learnable polynomial convolution kernel empowers the generalized Gram kernel with the capability of approximating the optimal filter function tailored to the specific scenario. A graph optimization target function and a pair-wise training objective are jointly employed to optimize the model parameters. Comparative experiments against a variety of state-of-the-art CF methods conclusively demonstrate the superiority of the proposed \method{}. In summary, the principal contributions of this work can be summarized and listed as follows:
\begin{itemize}[leftmargin=*]
    \item We conduct a comprehensive analysis of the expressiveness potential of existing node embedding-based models and introduce \method{}, a state-of-the-art graph filter-based method capable of approximating the optimal response function for CF problems.
    \item We propose the generalized normalization of Gram matrices, empowering the model to capture various spectral characteristics using a set of generalized Gram filters.
    \item Extensive experiments on three real-world recommendation datasets validate the effectiveness of \method{}. Further studies and analysis reveal the functionality of \method{} and the generalization capability of the model parameters.
\end{itemize}

\section{Related Works}

\subsection{Collaborative Filtering}
Collaborative Filtering (CF) tasks aim to recommend a personalized set of items to users based on similarities derived from their past interaction history \cite{goldberg1992using}. Traditional CF methods typically frame CF as a matrix factorization problem \cite{mnih2007probabilistic,koren2009matrix}, seeking to extract the most informative components from the original interaction matrix. In contrast, more recent ranking-based approaches have concentrated on minimizing ranking loss between selected positive and negative items \cite{rendle2012bpr,tay2018latent,mao2021simplex}. Another widely explored class of methods focuses on modeling the transition relationships between items \cite{brin1998pagerank,gori2007itemrank,he2016birank} with random processes such as Markov Chain.

The growing interest in graph neural networks and graph-based methodologies has led recent research in recommender systems to widely adopt graph-based models, such as sequence recommendation \cite{wu2019session,qiu2020gag,qin2023learning}, social networks \cite{fan2019graph,fan2019deep}, and location-based recommendation \cite{wang2022graph,qin2023disenpoi}. The message-passing structure of Graph Neural Networks (GNNs) empowers them to capture the topological structures and exploit implicit similarities present in multi-hop neighborhoods of nodes. Among the notable contributions in this area, NGCF \cite{wang2019neural} stands as a pioneering work that utilizes graph convolution networks to process user-item interaction bipartite graphs. LightGCN \cite{he2020lightgcn} refines the message function introduced by NGCF and has become a popular benchmark for graph-based CF models. UltraGCN \cite{mao2021ultragcn} delves into the nature of LightGCN's message function and enhances its expressiveness accordingly. CAGCN \cite{wang2023collaboration} introduces a recommendation-oriented topological metric to enhance link prediction accuracy. Furthermore, JGCF \cite{guo2023manipulating} extends the message function by incorporating polynomial filters to empower the model with the ability of manipulating graph signals, and has achieved promising results.

\subsection{Graph Signal Processing}
In addition to the traditional message-passing scheme employed by GNNs, there has been a significant focus on Graph Signal Processing (GSP) methods, which examine the filtering properties of graph operators from a spectral perspective \cite{ramakrishna2020user,dong2020graph,liu2022revisiting}. In a typical GSP model, input node features are treated as graph signals, and filters derived from the graph topology are applied to process these signals. Inspired by the pioneering work in graph convolution networks \cite{defferrard2016convolutional}, cutting-edge researches in the field of GNNs has unveiled the potential of high-order polynomial graph filters for expressing arbitrary smooth filter functions \cite{he2021bernnet,wang2022powerful}.

Inspired by the success of GSP, there have been attempts to integrate the capabilities of graph filters into CF models. EASE\textsuperscript{R} \cite{steck2019embarrassingly} studies the closed-form solution of the training objective of linear filters. Furthermore, GF-CF \cite{shen2021powerful} and PGSP \cite{liu2023personalized} propose to directly process signals from the rows of the interaction matrix and achieve promising results in various CF tasks. FIRE \cite{xia2022fire} proposes to incorporate graph filters into incremental recommendations. However, it's important to note that existing GSP methods still heavily rely on manually crafted filter structures, which do not fully harness the potential of graph filters.

\section{Preliminary}
In this section, we will provide a concise overview of graph-based Collaborative Filtering and introduce the notations used in graph signal processing for clarity in the following sections.

\subsection{Graph-based Collaborative Filtering}

A typical collaborative filtering task is conceptualized as a matrix completion problem that requires recovering a matrix with missing values. Specifically, the observed binary interaction matrix $R\in\{0,1\}^{m\times n}$ is derived from an interaction system involving $m$ users and $n$ items, where $R_{i,j}=1$ signifies the observation of an interaction between user $i$ and item $j$. The objective of the collaborative filtering task is to acquire the reconstructed matrix $R^* \in {0,1}^{m \times n}$ for augmenting $R$ with additional interactions that match similar users and items according to the collaborative affinity obtained from original matrix $R$.

For graph-based methods, the collaborative filtering task are further expanded to a link prediction task. Specifically, the corresponding user-item bipartite graph is formulated with the adjacency matrix $A=\begin{bmatrix}
    0 & R \\ R^T & 0
\end{bmatrix}$. The majority of graph-based methods adapt normalized graph convolution \cite{kipf2016semi} to propagate messages on the bipartite graph. Specifically, the normalized interaction matrix is calculated with a row- and column-wise normalization of $R$:
\begin{equation}
\tilde{R}=D_U^{-\frac{1}{2}}RD_I^{-\frac{1}{2}},  
\end{equation}
to balance the edge weights according to node degree. Similarly, the normalized adjacency matrix $\tilde{A}=\begin{bmatrix}
    0 & \tilde{R} \\ \tilde{R}^T & 0
\end{bmatrix}$. In practice, another commonly used graph is the item's Gram graph, represented by its adjacency matrix denoted as $\tilde{G}_I = \tilde{R}^T\tilde{R}$, which captures the collaborative relationships among items. Similarly the user's Gram matrix is formumlated as $\tilde{G}_U=\tilde{R}\tilde{R}^T$.

\subsection{Graph Signal Processing}

Given a weighted graph $\mathcal{G}=\{\mathcal{V}, \mathcal{E}\}$, where $\mathcal{V}$ represents the node set and $\mathcal{E}$ represents the edge set, the Laplacian matrix $L$ of $\mathcal{G}$ is defined with $L=D-A$, where $D=diag\{d_1,...,d_n\}$ is diagonal degree matrix. Specifically, the graph signals is presented as $x\in\mathbb{R}^{|\mathcal{V}|}$, with each component $x_i$ representing the signal response at node $i$. Graph filters are defined as mappings on graph signals and depends on the topological structure of $\mathcal{G}$.
% Graph filters are defined as mappings on graph signals, denoted as $f_{\mathcal{G}}:\mathbb{R}^{|\mathcal{V}|}\rightarrow \mathbb{R}^{|\mathcal{V}|}$ and the function $f_{\mathcal{G}}$ depends on the topological structure of $\mathcal{G}$.

To characterize the smoothness of graph signals, we formulate the graph derivative of a given signal $x$ using the graph quadratic form \cite{shuman2013emerging}, formulated as follows:
\begin{equation}
    S_\mathcal{G}(x)=x^TLx.
\end{equation}
The smoothness of graph signals on a given graph structure reflects the overall differences between adjacent nodes.

As the Laplacian matrix $L$ is positive semi-definite, it can be further factorized through its eigendecomposition: $L=U^T\Lambda U$. Where $\Lambda=diag\{\lambda_1,...,\lambda_{|\mathcal{V}|}\}$ is a diagonal matrix that composed of eigenvalues of $L$, $U=[u_1,...,u_{|\mathcal{V}|}]$ represents a set of corresponding normalized orthogonal eigenvectors. Graph filters are typically constructed based on the Graph Fourier Transform (GFT) $\hat{x}=Ux$, which maps the graph signal $x$ into the graph eigenspace:
\begin{equation}
    f_{\mathcal{G}}(x)=U^Tdiag\{f(\lambda_1),...,f(\lambda_n)\}Ux.
\end{equation}

\section{Methodology}

\begin{figure*}
\centering

\includegraphics[width=0.9\linewidth]{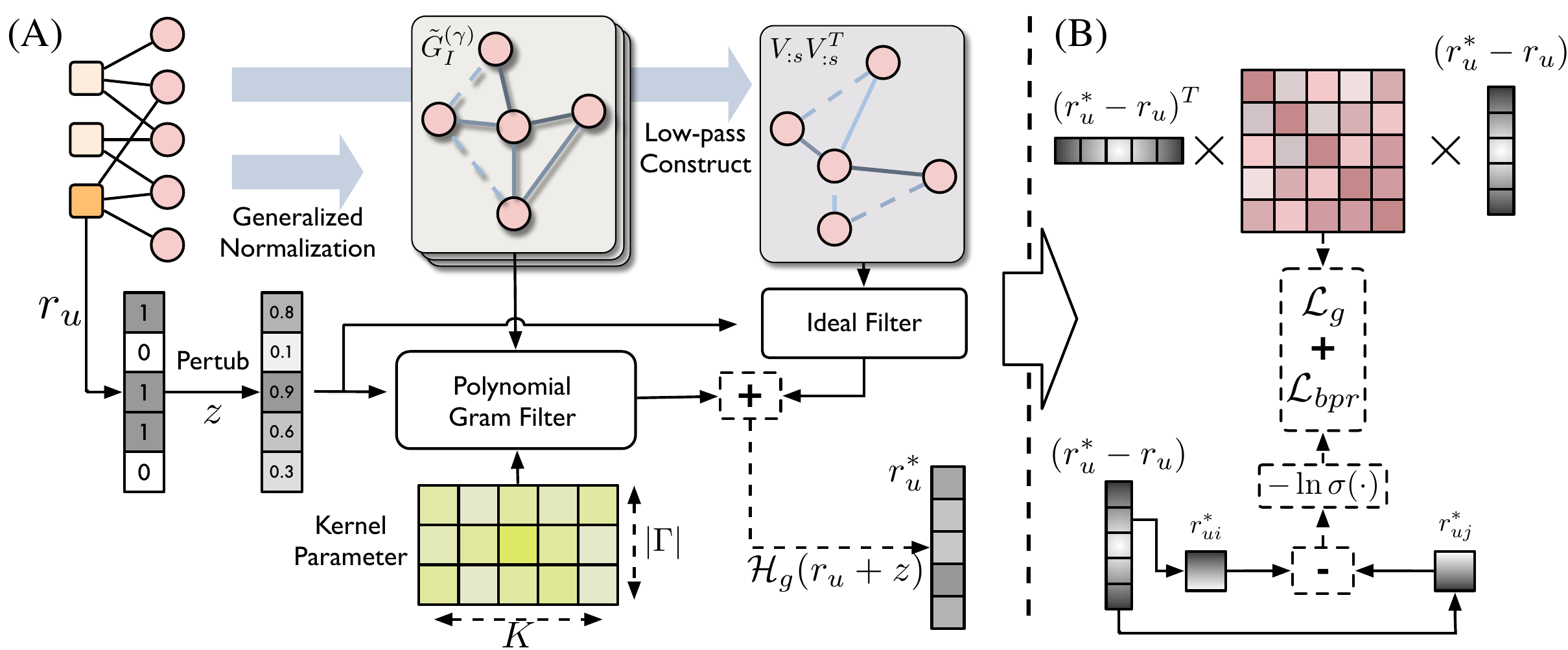}

\caption{A general illustration of the training framework of \method{}.} 
\label{fig:main}
\end{figure*}

In this section, we will start with conducting a brief review of the current node embedding-based graph methodologies and discuss their capacity limits. After that, we will introduce the proposed approach, namely \textit{\method{}}. As illustrated in Figure \ref{fig:main}, \method{} employs a multi-channel graph convolution operator, that comprises a series of generalized Gram convolution kernels. Finally, we will elucidate the optimization methodology employed to acquire the optimal polynomial approximation graph filters for collaborative filtering.

\subsection{From Embedding to Filter-based Models}
\label{sec:expressiveness}
\begin{figure}
\centering
\includegraphics[width=\linewidth]{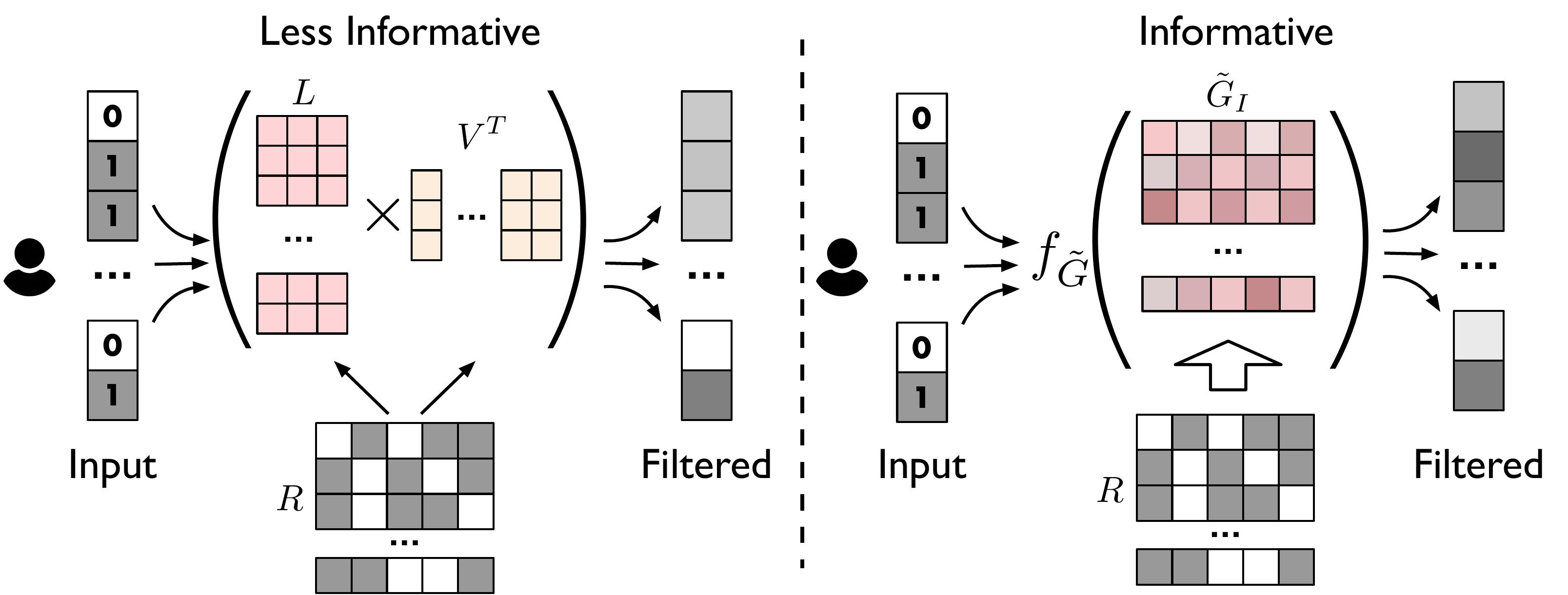}
\caption{Comparison between node embedding-based method (left) and graph filter-based methods (right).}
\label{fig:filter}
\end{figure}

Numerous graph-based Collaborative Filtering methods have emerged, employing node embeddings to represent users and items, and subsequently, to compute recommendation scores. Nevertheless, the full range of expressive capabilities and constraints associated with these node embedding-based methods remains to be explored. As an illustration, we consider the polynomial graph convolution model and dive into its capacity of recovering missing interactions. An illustrative comparison between two types of methods will be discussed and is presented in Figure \ref{fig:filter}.

To formulate a standard embedding-based method, like LightGCN \cite{he2020lightgcn} and JGCF \cite{guo2023manipulating}, it is parameterized by an embedding matrix $E=\begin{bmatrix}
    E_U,E_I
\end{bmatrix}\in\mathbb{R}^{(m+n)\times d}$, where $E_U,E_I$ represent node embeddings for users and items respectively. The propagation process can then be expressed as follows:
\begin{equation}
P(\tilde{A})E=\sum\limits_{k=0}^K\alpha_k\Tilde{A}^kE,
\end{equation}
where $\alpha_k$s denotes the coefficients of the polynomial graph filter. More specifically, the reconstructed interaction matrix is derived from the inner-product between the propagated embeddings of users and items:
\begin{equation}
\label{eq:score}
    R^*=[P(\tilde{A})E]_{:m}\cdot [P(\tilde{A})E]_{m+1:}^T.
\end{equation}

To explore the expressiveness of the resulting reconstructed matrix in Equation \ref{eq:score}, we conduct a more in-depth analysis of the node embeddings obtained through the graph convolution layers.
\begin{theorem}
\label{th:factorization}
    For any polynomial graph filter $P(\tilde{A})$, there exists $L,V\in\mathbb{R}^{n\times d}$ that satisfy the relationship:
    \begin{equation}
        R^*=\tilde{R}\cdot LV^T,
    \end{equation}
    where both $L$ and $V$ are linear combinations of $E_I$ and $\tilde{R}E_U$. The computation of these two matrices depends solely on the polynomial coefficients $\alpha_k$ and the item Gram matrix $\tilde{G}_I$.
\end{theorem}

As demonstrated in Theorem \ref{th:factorization}, the node embedding-based approach can be regarded as a specific type of factorization for the observed interaction matrix $\tilde{R}$. When $P(\tilde{A})=I$, the model degenerates to a low-rank factorization \cite{koren2009matrix} of the interaction matrix. Consequently, we draw two crucial insights from Theorem \ref{th:factorization}:
\begin{itemize}[leftmargin=*]
    \item The expressiveness of node embedding-based graph methods is constrained by the embedding size $d$. This limitation arises because the rank of the reconstructed matrix $R^*$ is bounded by $\min\{rank(U), rank(V)\} \leq d$. Consequently, their capacity to recover missing interactions is fundamentally capped by the predefined embedding size $d$.
    \item The normalized item Gram matrix $\tilde{G}_I$ is the key to overcome the limitations due to the node embedding size. This significance arises from the fact that the minimum rank of $\tilde{G}_I$ can be bounded using the Sylvester inequality: 
    \begin{equation}
        rank(\tilde{G}_I)\ge 2rank(\tilde{R})-m\approx m.
    \end{equation}
    In practice, the number of users $u$ is often significantly greater than feasible embedding size $d$. This observation suggests that replacing the factorization involving $L$ and $V$ with transformations based on $\tilde{G}_I$ can unlock a wealth of additional information and benefited from the higher rank of the transformation.
    % , which indicates replacing the factorization $L$ and $V$ with transformations of $\tilde{G}_I$ could bring more information out of higher rank of the transformation matrix.
    % \item The normalized item Gram matrix $\tilde{G}_I$ plays a pivotal role in graph-based methods. This is due to the fact that the entire propagation process can be accurately represented using $\tilde{G}_I$ in combination with a set of coefficients.
\end{itemize}

Motivated by the insights gleaned from the aforementioned observations, we introduce the proposed graph filter-based \method{}, which doesn't require node embeddings and is built upon stacking layers of the generalized normalization of Gram convolution kernel.

\subsection{Generalized Gram Convolution}

Recall that the normalized item Gram matrix is derived from the product of the normalized interactions, given by $\tilde{G}_I=\tilde{R}^T\tilde{R}\in\mathbb{R}^{n\times n}$. This particular form of the Gram matrix has been widely adopted to design graph filters in prior studies \cite{steck2019embarrassingly,shen2021powerful,liu2023personalized} for capturing item co-occurrence patterns. Despite its proven effectiveness in practical applications, its characteristics as a graph filter have remained unexplored. Therefore, we undertake a comprehensive analysis from a graph spectral perspective to develop a more potent polynomial Gram filter for our proposed \method{}. 

\subsubsection{Generalized Gram Filter}

In previous works, the Laplacian matrices are often symmetrically normalized, which constrained the expressiveness of the associated graph filter. Studies on graph convolution \cite{gasteiger2018predict} have indicated that employing random-walk normalized matrices, akin to PageRank \cite{brin1998anatomy} could yield more personalized propagation outcomes. We extend the idea to encompass more flexible situations and propose the notion of \textit{Generalized Normalization}. For any given adjacency matrix $A$, the generalized normalization of its Laplacian matrix  is defined as follows:
\begin{equation}
\tilde{L}^{(\gamma)}=D^{-\gamma}(D-A)D^{\gamma-1}=I-D^{-\gamma}AD^{\gamma-1}.
\end{equation}
When $\gamma=\frac{1}{2}$, this generalized normalization is equivalent to symmetrical normalization, and when $\gamma=1$ it becomes random-walk normalization. Similarly, the generalized normalization of gram matrix $G_I$ is formulated as:
\begin{equation}
\label{eq:normalize}
    \tilde{G_I}^{(\gamma)}=\tilde{R}^{(-\gamma)T}\tilde{R}^{(-\gamma)}=D_I^{-\gamma}R^TD_U^{-1}RD_I^{\gamma-1}.
\end{equation}

By introducing the concept of generalized normalization for $G_I$, we can leverage its adaptable spectral structure.
\begin{theorem}
\label{th:generalize}
    For any $\gamma\in[0,1]$, the Laplacian of $\tilde{G}_I^{(\gamma)}$ shares the same set of eigenvalues that satisfies the following inequality:
    \begin{equation}
        0\le\lambda_1\le...\le\lambda_n\le1.
    \end{equation}
    Moreover, for any $\mu_i^{(\gamma_1)}$ and $\mu_i^{(\gamma_2)}$ are eigenvectors associated with $\lambda_i$ for $\tilde{G}_I^{(\gamma_1)}$ and $\tilde{G}_I^{(\gamma_2)}$ respectively, the relationship holds:
    \begin{equation}
        \mu_i^{(\gamma_1)}=D_I^{\gamma_1-\gamma_2}\mu_i^{(\gamma_2)}
    \end{equation}
\end{theorem}
Theorem \ref{th:generalize} implies that Gram matrices normalized with different values of $\gamma$ share similar spectral spaces, while still maintaining their distinct eigenspaces. This property of generalized normalization helps us to build Gram filters capable of capturing spectral features from various perspectives.

\subsubsection{Polynomial Graph Convolution}
Having obtained the generalized Gram filters $\tilde{G}_I^{(\beta)}$, our next step is to devise a polynomial spectral filter based on them. To formulate, consider an arbitrary normalized Laplacian matrix $\tilde{L}$ with its eigendecomposition $\tilde{L}=U^{T}\Lambda U$. Our objective is to identify an optimal linear filter $h(\lambda)$ which depicts the transition between observed interactions and reconstructed interactions. More specifically, for a user with its interaction history $r_u\in\{0,1\}^{n}$, the Gram filter is applied as:
\begin{equation}
    r_u^*=h(\tilde{L})r_u=U^Th(\Lambda)Ur_u=U^Tdiag\{h(\lambda_1),...,h(\lambda_n)\}Ur_u.
\end{equation}

Inspired by previous research on polynomial graph convolution \cite{defferrard2016convolutional,he2021bernnet,wang2022powerful}, we express $h(\lambda)$ as a linear combination of a set of orthogonal polynomial basis functions:
\begin{equation}
\label{eq:single_filter}
    h(\lambda)=\sum\limits_{k=0}^K\theta_k\cdot P_k(1-\lambda),
\end{equation}
where $\theta_k$ represents the linear coefficient and $P_k(\lambda)$ denotes the $k$-th basis function, which incorporates a $k$-order polynomial of $\lambda$. Equation \ref{eq:single_filter} defines a $K$-order filter function that can be applied to the spectral domain of the Gram item graph using $h(\tilde{G})$. There exist various options for selecting the polynomial basis, such as the Chebyshev basis and the Bernstein basis.

The derived generalized normalization of Gram matrix in Equation \ref{eq:normalize} enables the model to conduct graph filtering on an input signal using filters possessing distinct eigenspace structures. This capability empowers the resulting polynomial filter to leverage the rich spectral features of graph signals. In summary, the proposed generalized Gram convolution kernel is defined as:
\begin{equation}
\label{eq:filter_kernel}
    \mathcal{H}_{P}({G}_I)=\frac{1}{|\Gamma|}\sum\limits_{\gamma\in\Gamma}\sum\limits_{k=0}^K\theta_k^{(\gamma)}P_k(\tilde{G}_I^{(\gamma)}),
\end{equation}
where $\theta_k^{(\gamma)}$ is the parameter of the generalized convolution kernel, which can be optimized using gradient-based methods. By aggregating the filtering outcomes from different eigenspaces, the generalized Gram convolution is capable of learning more comprehensive filter functions to recover missing interactions.

\subsubsection{Low-pass Enhancement}
Despite the generalized Gram convolution kernel presented in Equation \ref{eq:filter_kernel} offers a polynomial approximation for the desired graph filter, the optimal filter describing the transformation between $R$ and $R^*$ may not exhibit the desired smoothness when expressed as a linear filter. As suggested in prior studies \cite{shen2021powerful,cai2023lightgcl}, a commonly adopted approach is to employ the ideal low-pass filter to enhance the presence of low-frequency components in the input interaction signals.

 To formulate, the ideal $s$-pass filter is obtained from the Singular Value Decomposition (SVD) on the Gram matrix, written as:
 \begin{align}
    \tilde{G}_I&=U\Sigma V^T \\
    \mathcal{H}_s({G}_I)&=V_{:s}V_{:s}^T,
 \end{align}
 where the decomposed Gram matrix $\tilde{G}_I$ could be generally normalized to any order $\beta$and we default to $\beta=\frac{1}{2}$. The filter function for $\mathcal{H}_s$ is determined by the low-pass parameter $s$:
 \begin{equation}
     h(\lambda)=\begin{cases}
         1 & \lambda\le\lambda_s \\
         0 & otherwise.
     \end{cases}
 \end{equation}

The ideal low-pass filter enhances the input interaction signals by eliminating high-frequency noise and retaining the more informative low-frequency components.

\subsubsection{Comprehensive Polynomial Filter}

So far we have developed two key filters: the polynomial generalized Gram filter $\mathcal{H}_P$ and the ideal low-pass filter $\mathcal{H}_s$. The overall filtering process for the \method{} can be expressed as follows:
\begin{equation}
\label{eq:inference}
    \mathcal{H}_gr_u=(\mathcal{H}_P+\omega\mathcal{H}_s)r_u,
\end{equation}
where the hyperparameter $\omega$ is for balancing the trade-off between low-pass signal enhancement and polynomial approximation.

In practice, to reduce the computational complexity  associated with obtaining the filtered signal in Equation \ref{eq:inference}, we adopt the factorized form of the comprehensive filters that fully leverage sparse matrix multiplication:
\begin{equation}
\begin{aligned}
\label{eq:compute}
\mathcal{H}_gr_u&=\mathcal{H}_Pr_u+\omega\mathcal{H}_sr_u \\
&=\frac{1}{|\Gamma|}\sum\limits_{\gamma\in\Gamma}\sum\limits_{k=0}^K\theta_k^{(\gamma)}P_k(D_I^{-\gamma}R^TD_U^{-1})(RD_I^{\gamma-1}r_u)+\omega V_{:s}(V_{:s}^Tr_u).
\end{aligned}
\end{equation}
 
\subsection{Model Optimization}

In the previous sections, we have outlined the filtering process of the polynomial Gram filter. Nonetheless, it still needs a target function to optimize the convolution kernel of $\mathcal{H}_P$. Consequently, we introduce the optimization objective of \method{} to learn the optimal approximation of the desired graph filter.

\subsubsection{Graph Optimization Objective}

To optimize $\mathcal{H}_P$ as a low-pass filter, we formulate the graph optimization objective with a graph optimization problem as follows:
\begin{equation}
\label{eq:target_graph}
    \mathcal{L}_g=(r_u^*-r_u)^T(I-\tilde{G}_I^{(\frac{1}{2})})(r_u^*-r_u), %+ \lVert r_u^*-r_u\rVert_2^2,
\end{equation}
where the filtered $r_u^*$ is obtained through:
\begin{equation}
    r_u^*=\mathcal{H}_g(r_u+{z}),\ {z}\sim\mathcal{N}(0,\epsilon I).
\end{equation}
The hyperparameter $\epsilon$ is used to control the level of disturbance added to the original interaction signal of $u$.

By introducing random noise $z$ to the input signal, the graph optimization objective in Equation \ref{eq:target_graph} encourages the learned convolution kernel to behave as a robust filter against high-frequency noise in signals while preserving valuable low-frequency components for recommendation.

\subsubsection{Bayesian Ranking Optimization}

In addition to optimizing \method{} as a low-pass graph filter, we incorporate Bayesian Ranking Loss (BPR) \cite{rendle2012bpr} to enhance the ranking performance of the learned graph filter. The BPR objective function is formulated as follows:
\begin{equation}
    \mathcal{L}_{bpr}=\sum\limits_{(u,i,j)\in\mathcal{O}}-\ln\sigma(r_{ui}^*-r_{uj}^*),
\end{equation}
where $\mathcal{O}=\{(u,i,j)|R_{u,i}=1,R_{u,j}=0\}$ denotes the set of sampled data pairs. In each iteration, we randomly select negative items for each observed user-item interaction pair.

In summary, we combine the two target function to formulate the optimization objective of \method{}:
\begin{equation}
    \mathcal{L}=\mathcal{L}_g+\mathcal{L}_{bpr}.
\end{equation}

\section{Experiment}
\begin{table}
\centering
\caption{Descriptive statistics of the used datasets.}
\label{tab:statics}
\setlength{\tabcolsep}{3pt}
\begin{tabular}{ccccc} 
\toprule %\hline
\textbf{Dataset} & \#User & \#Item & \#Interactions & Density \\
\midrule
Amazon-Book & 52,643 & 91,599 & 2,984,108 & 0.062\% \\
Yelp2018 & 31,668 & 38,048 & 1,561,406 & 0.130\% \\
Gowalla & 29,858 & 40,981 & 1,027,370 & 0.084\% \\
\bottomrule %\hline
\end{tabular}
\end{table}

\begin{table*}
\centering
\caption{The test results of \method{} and all baseline methods. The highest performance is emphasized with bold font and the second highest is marked with underlines.}
% \textbf{$\star$} indicates that \method{} outperforms the best baseline model at a p-value<0.05 level of unpaired t-test.}
\label{tab:overall} 
% \begin{tabularx}{\linewidth}{cc YYYYYY}
\begin{tabular}{cc cccccc}
\toprule %\hline
\multirow{2}{*}{ } & \multirow{2}{*}{Method} & \multicolumn{2}{c}{\textbf{Amazon-Book}} & \multicolumn{2}{c}{\textbf{Yelp2018}} & \multicolumn{2}{c}{\textbf{Gowalla}} \\ 
\cmidrule[0.5pt](lr){3-4}\cmidrule[0.5pt](lr){5-6}\cmidrule[0.5pt](lr){7-8}
& & Recall@20 & NDCG@20 & Recall@20 & NDCG@20 & Recall@20 & NDCG@20 \\
\midrule 

\multirow{5}{*}{\shortstack{\textbf{CF-based}\\\textbf{Models}}} 
& BPR \cite{rendle2012bpr} & 0.0250 & 0.0196 & 0.0433 & 0.0354 & 0.1291 &  0.1109 \\
& NeuMF \cite{he2017neural} & 0.0258 &  0.0200 & 0.0451 & 0.0363 & 0.1399 &  0.1212 \\
& ENMF \cite{chen2020efficient} & 0.0359 & 0.0281 & 0.0624 & 0.0515 & 0.1523 & 0.1315 \\
& YoutubeDNN \cite{covington2016deep} & 0.0502 & 0.0388 & 0.0686 &  0.0567 & 0.1754 & 0.1473 \\
& SimpleX \cite{mao2021simplex} & 0.0583 & 0.0468 & 0.0701 &  0.0575 & 0.1872 & 0.1557 \\
\midrule

\multirow{5}{*}{\shortstack{\textbf{GNN-based}\\\textbf{Models}}}
& APPNP \cite{gasteiger2018predict}& 0.0384 & 0.0299 &  0.0635 & 0.0521 & 0.1708 &  0.1462 \\
& LightGCN \cite{he2020lightgcn} & 0.0411 &  0.0315 & 0.0649 & 0.0530 & 0.1830 & 0.1554 \\
& DGCF \cite{wang2020disentangled} & 0.0422 & 0.0324  & 0.0654 & 0.0534 & 0.1842 &  0.1561 \\
& UltraGCN \cite{mao2021ultragcn} & 0.0681 & 0.0556 & 0.0683 & 0.0561 & 0.1862 & 0.1580 \\
& CAGCN\textsuperscript{*} \cite{wang2023collaboration} & 0.0510 & 0.0403 & 0.0708 & \underline{0.0586} & 0.1878 & 0.1591 \\
& JGCF \cite{guo2023manipulating} & 0.0692 & 0.0559 & 0.0701 & 0.0579 & 0.1894 & 0.1593 \\
\midrule

\multirow{4}{*}{\shortstack{\textbf{Graph Filter}\\\textbf{Models}}} 
& EASE\textsuperscript{R} \cite{steck2019embarrassingly} & 0.0710 & 0.0567 & 0.0657 & 0.0552 & 0.1765 & 0.1467 \\
& GF-CF \cite{shen2021powerful} & 0.0710 & 0.0584 & 0.0697 & 0.0571 & 0.1849 & 0.1518 \\
& LinkProp-Multi \cite{fu2022revisiting} & \underline{0.0721} & \underline{0.0588} & 0.0690 & 0.0571 & 0.1908 & 0.1573 \\
& PGSP \cite{liu2023personalized} & 0.0712 & 0.0587 & \underline{0.0710} & {0.0583} & \underline{0.1916}& \underline{0.1605} \\
\midrule

% & \method{}\textsubscript{single} & - & - & - & - & - & - \\
% & \method{}\textsubscript{linear} & - & - & - & - & - & - \\
& \method{} & \textbf{0.0730} & \textbf{0.0605} & \textbf{0.0715} & \textbf{0.0590} & \textbf{0.1934} & \textbf{0.1621}  \\
& Improv. & 1.5\% & 2.9\% & 0.71\% & 0.68\% & 0.94\% & 1.00\% \\
\bottomrule % \hline
\end{tabular}
% \end{tabularx}
\end{table*}

In this section, we present a series of comprehensive experiments conducted on three collaborative filtering datasets. These experiments aim to showcase the effectiveness and robustness of the proposed \method{}. Additionally, we perform in-depth ablation and parameter studies to investigate the functionality of different modules within \method{} and analyze its sensitivity to hyperparameters.

\subsection{Experiment Settings}
\subsubsection{Dataset and Evaluation Metric}
We evaluate the performance of \method{} and compared baseline methods on three widely adopted datasets, namely \textbf{Amazon-Book}, \textbf{Yelp2018}, and \textbf{Gowalla}. The statistics of the used datasets are in Table \ref{tab:statics}. We adopt the same train/test set split to keep consistency with previous works \cite{he2020lightgcn,shen2021powerful,mao2021simplex}. We randomly choose 10\% of the train set to tune hyperparameters for our model.

To evaluate the performance of our \method{}, we adopt two widely used evaluation metrics, namely Recall@K and NDCG@K. We set the value of K to 20 for consistency across different methods.

\subsubsection{Compared Baselines}

We compare the performance of the proposed \method{} with a wide range of baseline models from three perspectives, listed as follows:
\begin{itemize}[leftmargin=*]
    \item \textbf{CF-based Models}: Traditional collaborative filer methods that focus on depicting the collaborative affinity between users and items. Include models that are based on matrix factorization, i.e. NeuMF \cite{he2017neural} and ENMF \cite{chen2020efficient}, or based on two-tower structures , i.e. BPR \cite{rendle2012bpr}, YoutubeDNN \cite{covington2016deep} and SimpleX \cite{mao2021simplex}.
    \item \textbf{GNN-based Models}: Models that leverage GNNs to capture the high-order similarities between nodes on interaction graphs. Including APPNP \cite{gasteiger2018predict}, LightGCN \cite{he2020lightgcn}, DGCF \cite{wang2020disentangled}, UltraGCN \cite{mao2021ultragcn}, CAGCN \cite{wang2023collaboration} and JGCF \cite{guo2023manipulating}.
    \item \textbf{Graph Filter Models}: Models that integrate graph filters to processing interaction graph signals. Including EASE\textsuperscript{R} \cite{steck2019embarrassingly}, GF-CF \cite{shen2021powerful}, LinkProp-Multi \cite{fu2022revisiting} and PGSP \cite{liu2023personalized}.
\end{itemize}

\subsubsection{Implementation Details}
We have implemented the proposed \method{} using the PyTorch framework. For the baseline methods we compare against, we either duplicate the performance results from their original papers or reproduce the results based on the open-source implementations. In the case of our \method{}, the normalization order set $\mathcal{B}$ consists of 4 values chosen from the interval $[0,1]$ with a fixed step size of 0.1, and the polynomial order is fixed at $K=5$. We apply a dropout method on the convolution kernel with dropout rates tuned from $\{0.2, 0.4, 0.6, 0.8\}$. The convolution parameters are optimized using stochastic gradient descent with a learning rate of $lr=10^{-3}$. Various polynomial basis functions could be employed in Equation \ref{eq:single_filter}, we have opted to select $P_k$ from the following basis functions: Monomial basis, Chebyshev basis, Bernstein basis, Jacobi basis, and Hermite basis.

\subsection{General Comparison}

We conduct the general experiments of the proposed \method{} and the compared methods on the aforementioned three datasets and record their recommendation performance. From the results reported in Table \ref{tab:overall}, we make the following observations:
\begin{itemize}[leftmargin=*]
    \item The proposed \method{} outperforms currently advanced methods, achieving state-of-the-art recommendation performance across all datasets. Specifically, \method{} outperforms the best baseline method by over 1.5\%, 0.71\%, and 0.94\% relative improvement on Recall@20, over 2.9\%, 0.68\%, and 1.00\% relative improvement on NDCG@20. This underscores the effectiveness of the idea of integrating generalized polynomial Gram filters and ideal low-pass filters in recovering missing interactions.
    \item Graph-based methods exhibit superior performance compared to traditional CF methods, highlighting the advantages of leveraging graph structures to model collaborative affinity between users and items. Notably, graph filter-based methods consistently demonstrate superior and more stable performance in comparison to graph neural network (GNN)-based models. This observation aligns with the insights derived from Theorem \ref{th:generalize}.
    \item Graph filter-based methods tend to excel on sparser datasets, such as Amazon-Book and Gowalla, while other embedding-based methods exhibit better performance on relatively denser datasets like Yelp2018. This pattern suggests that models derive significant benefits from filter results that capture global relationships within interaction graphs, particularly in cases where training data is sparse.
\end{itemize}

\subsection{Ablation Study}
\begin{figure}
\centering
\begin{subfigure}{0.49\linewidth}
    \includegraphics[width=\linewidth]{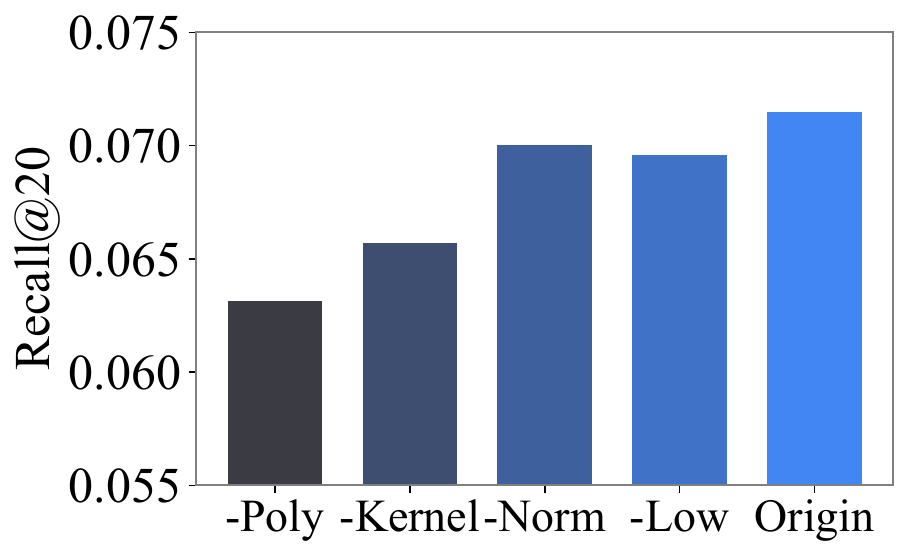}
    \caption{Recall on Yelp2018}
\end{subfigure}
\begin{subfigure}{0.49\linewidth}
    \includegraphics[width=\linewidth]{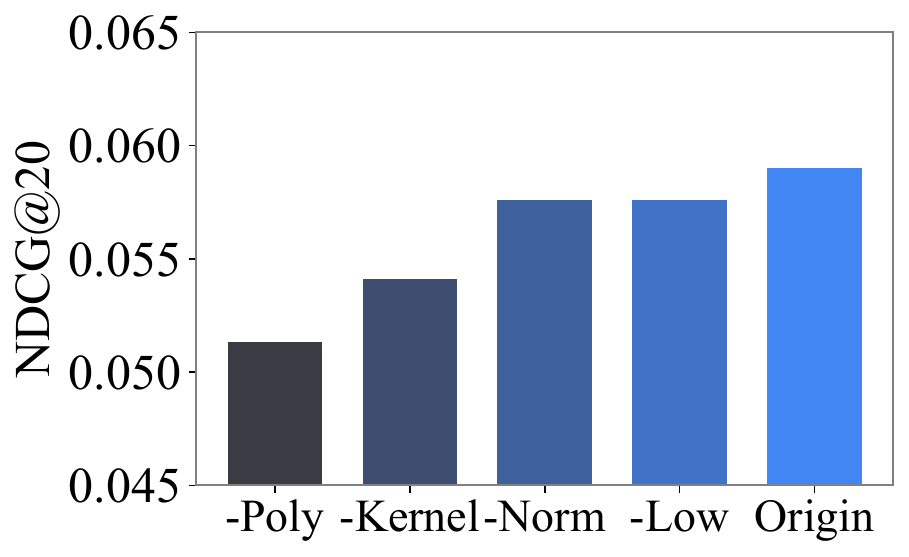}
    \caption{NDCG on Yelp2018}
\end{subfigure}

\begin{subfigure}{0.49\linewidth}
    \includegraphics[width=\linewidth]{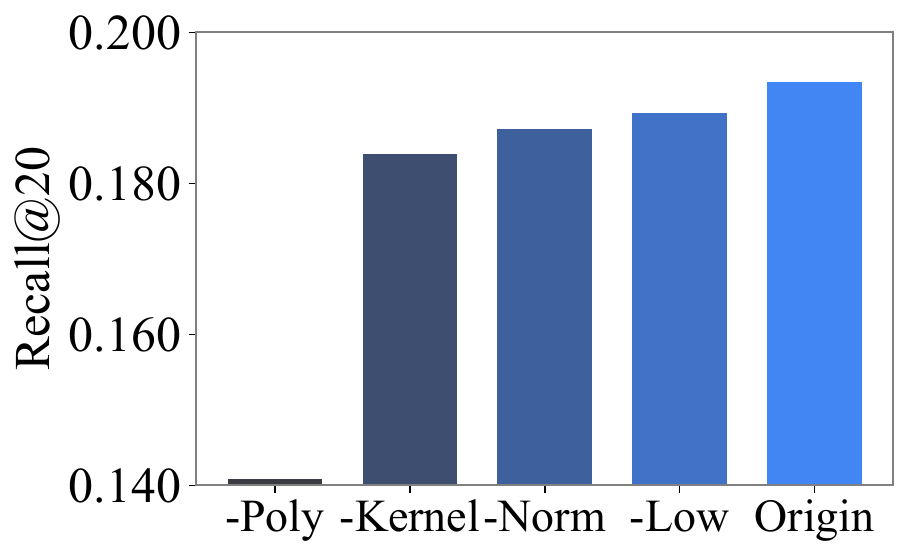}
    \caption{Recall on Gowalla}
\end{subfigure}
\begin{subfigure}{0.49\linewidth}
    \includegraphics[width=\linewidth]{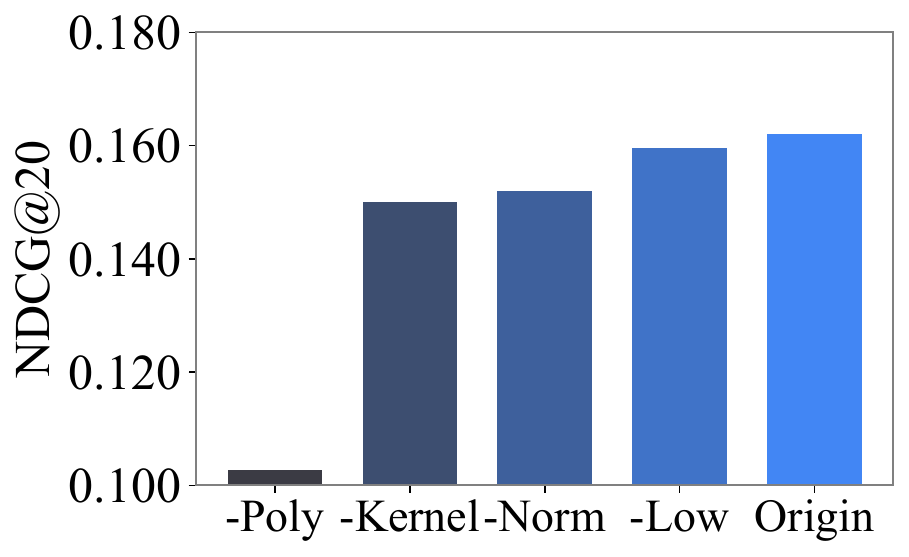}
    \caption{NDCG on Gowalla}
\end{subfigure}

\caption{Model performance with different settings.} 
\label{fig:ablation}
\end{figure}

The proposed \method{} incorporates a polynomial generalized Gram filter and an ideal low-pass filter to approximate the optimal filter function for recovering missing interactions. Additionally, a graph optimization objective function and a BPR ranking loss are leveraged to optimize the parameters of the polynomial filter. To gain a deeper understanding of the contributions of each module and how \method{} benefits from the proposed optimization components, we conducted comprehensive ablation studies. These studies help dissect the role and impact of each element within \method{}. %'s overall framework.

\subsubsection{Functionality of Graph Filters}

The two graph filters applied in \method{} serve distinct purposes: the flexible polynomial generalized Gram filter aims to approximate the optimal response function, while the ideal low-pass filter enhances low-pass components, thereby improving the model's expressiveness. To comprehensively assess the contributions of these components, we conducted experiments comparing the performance of \method{} with four variants:
\begin{enumerate}[leftmargin=*]
    \item W/O-Poly: This variant of \method{} excludes the polynomial Gram filter, meaning the input signals are solely processed with the ideal low-pass filter.
    \item W/O-Kernel: In this variant, we replace the polynomial convolution kernel in \method{} with a fixed 1-order normalized Gram.
    \item W/O-Norm: This variant of \method{} removes the generally normalized Gram kernels, using only the polynomial of $\tilde{G}_I^{-\frac{1}{2}}$.
    \item W/O-Low: In this variant, we omit the ideal low-pass filter $H_s$.
\end{enumerate}

By comparing the performance of these variants, we can gain insights into the individual contributions of each module in \method{}. Based on the results of ablation study presented in Figure \ref{fig:ablation}, we can draw the following conclusions:
\begin{itemize}[leftmargin=*]
    \item All components of \method{} helps to improve model's capability of modeling the optimal graph filter. Specifically, when the polynomial filter is removed, \method{} experiences the most significant performance decline, due to the removed relative signals. This observation indicates that the flexibility brought by a generalized polynomial Graph filter is the key element for \method{} to achieve its state-of-the-art performance.
    \item Among the other components, the absence of polynomial kernel in \method{} leads to the most notable performance decline. This suggests that the presence of polynomial Gram filters empowers the model to effectively capture informative features and approximate the optimal filter function.
\end{itemize}

\subsubsection{Effectiveness of Optimization Target}

\begin{table}
\centering
\caption{Performance of different optimization targets.}
\label{tab:ablation} 
\begin{tabularx}{\linewidth}{c YYYYYY}
\toprule %\hline
\multirow{2}{*}{Method} & \multicolumn{2}{c}{\textbf{Amazon-Book}} & \multicolumn{2}{c}{\textbf{Yelp2018}} & \multicolumn{2}{c}{\textbf{Gowalla}} \\ 
\cmidrule[0.5pt](lr){2-3}\cmidrule[0.5pt](lr){4-5}\cmidrule[0.5pt](lr){6-7}
& R@20 & N@20 & R@20 & N@20 & R@20 & N@20 \\
\midrule

GF-CF & 0.0710 & 0.0584 & 0.0697 & 0.0571 & 0.1849 & 0.1518 \\
W/O-$L_{bpr}$ & 0.0718 & 0.0592 & 0.0709 & 0.0583 & 0.1920 & 0.1607 \\
W/O-$L_g$ \ \ \ & 0.0721 & 0.0600 & 0.0712 & 0.0587 & 0.1912 & 0.1599 \\
Origin & {0.0730} & {0.0605} & {0.0715} & {0.0590} & {0.1934} & {0.1621} \\
\bottomrule % \hline
% \end{tabular}
\end{tabularx}
\end{table}

\begin{figure}
\centering
\begin{subfigure}{0.49\linewidth}
    \includegraphics[width=\linewidth]{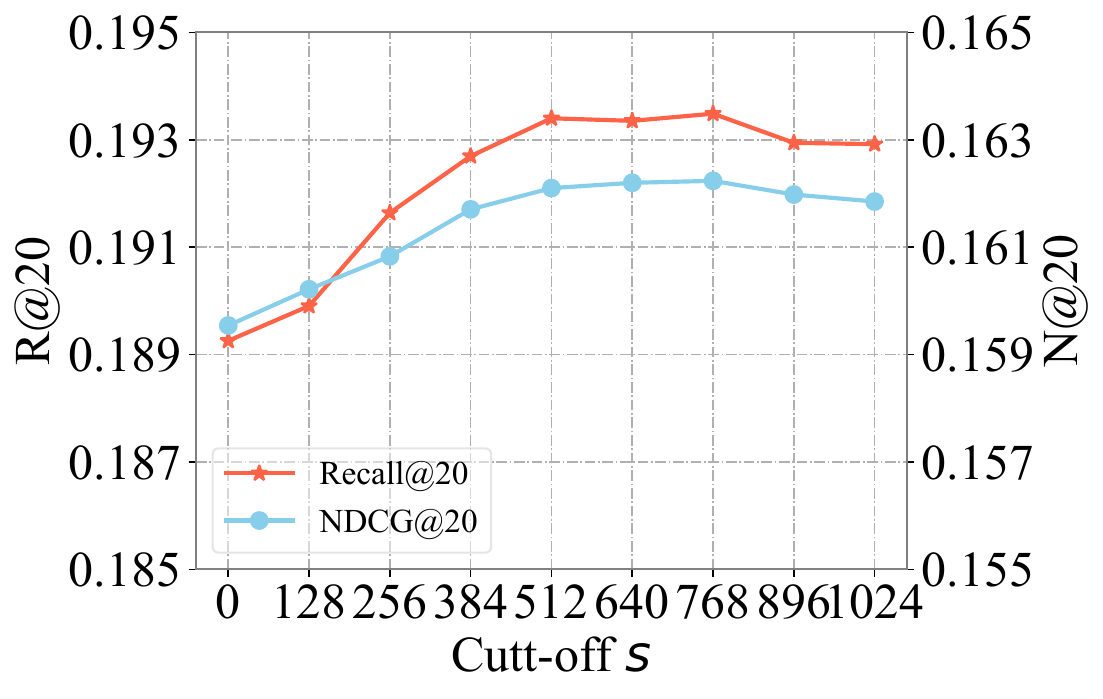}
    \caption{Results on Gowalla}
\end{subfigure}
\begin{subfigure}{0.49\linewidth}
    \includegraphics[width=\linewidth]{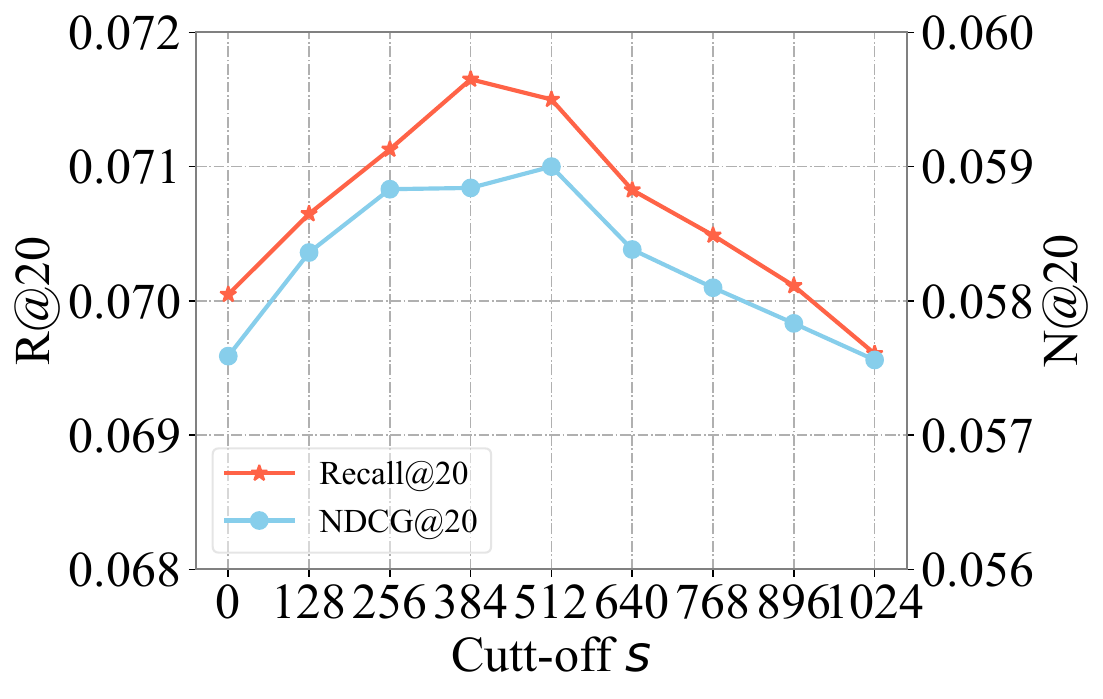}
    \caption{Results on Yelp2018}
\end{subfigure}

\caption{Model performance w.r.t. cut-off frequency $s$.} 
\label{fig:parameter}
\end{figure}

Recall that \method{} is optimized via a pair of optimization target: a graph-based optimization objective $\mathcal{L}_g$ and a Bayesian-based ranking loss $\mathbf{L}_{bpr}$. We conduct ablation studies on these two different optimization objectives to validate their effectiveness on \method{}'s performance. From the performance illustrated in Table \ref{tab:ablation}, we can observe that:
\begin{itemize}[leftmargin=*]
    \item Both $\mathcal{L}_g$ and $\mathcal{L}_{bpr}$ are crucial to maintaining stable and adaptable performance across different datasets. This observation indicates that the two objective functions complement each other: the graph optimization objective enhances \method{}'s robustness against high-frequency noise, while the pairwise ranking loss encourages \method{} to approximate the optimal filter for making personalized rankings and recommendations. 
    \item Different optimization objectives are more suitable for different scenarios. Specifically, $\mathcal{L}_g$ plays a more significant role in improving performance on the Gowalla dataset, whereas $\mathcal{L}_{bpr}$ appears to be the key to achieving better results on the Amazon-Books and Yelp2018 datasets. This indicates that the choice of optimization objective can be tailored to the characteristics of the dataset to enhance performance.
\end{itemize}

\subsection{Parameter Analysis}

The proposed \method{} involves pre-defined hyperparameters that can significantly influence recommendation performance. Therefore we conduct parameter studies to evaluate the sensitivity of \method{} to these hyperparameters.

The used ideal low-pass filter $\mathcal{H}_s$ depends on a cut-off frequency $s$ that defines the bandwidth of the filtered components. We conducted a parameter study by adjusting the value of $s$ and reporting the model's performance. From the results in Figure \ref{fig:parameter}  we can draw the following conclusions:
\begin{itemize}[leftmargin=*]
    \item The choice of cut-off frequency $s$ is critical to \method{}'s performance. A proper selection of $s$ empowers \method{} to effectively utilize the most informative part of each input interaction signal. In contrast, extremely large or small cut-off selection would lead to sub-optimal due to the missing components lying in the informative low-pass region of input signals.
    \item The optimal cut-off frequency $s$ varies across different datasets. This suggests that the most informative signal bandwidth is flexible and can adapt to the characteristics of different user-item interactions.
\end{itemize}
% Another important parameter is the weight $\omega$ of the low-pass signal, which balance different components of filtered signals. We conduct parameter studies on these two parameters simultaneously to analyse their collaborative influence. 

% From the results in Figure \ref{} we can conclude the choice of cut-off frequency $s$ and the low-pass weight $\omega$ are crucial to \method{}'s performance. Particularly, a proper cut-off $s$ empowers \method{} to make use of the most informative part of each input interaction signal. Additionally, appropriate $\omega$ is also important to \method{}, while a small $\omega$ leads to lack of low-pass information, a extremely large $\omega$ would make \method{} become rigid and lose its flexibility of the polynomial Gram filter.

\subsection{In-depth Study}

\subsubsection{Generalization of Convolution Kernel}

\begin{figure}
\centering
\begin{subfigure}{0.49\linewidth}
    \includegraphics[width=\linewidth]{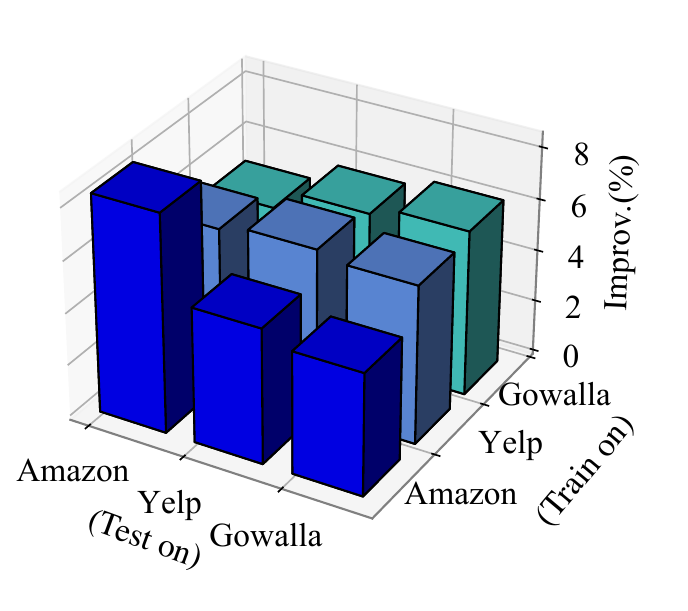}
    \caption{Recall@20}
\end{subfigure}
\begin{subfigure}{0.49\linewidth}
    \includegraphics[width=\linewidth]{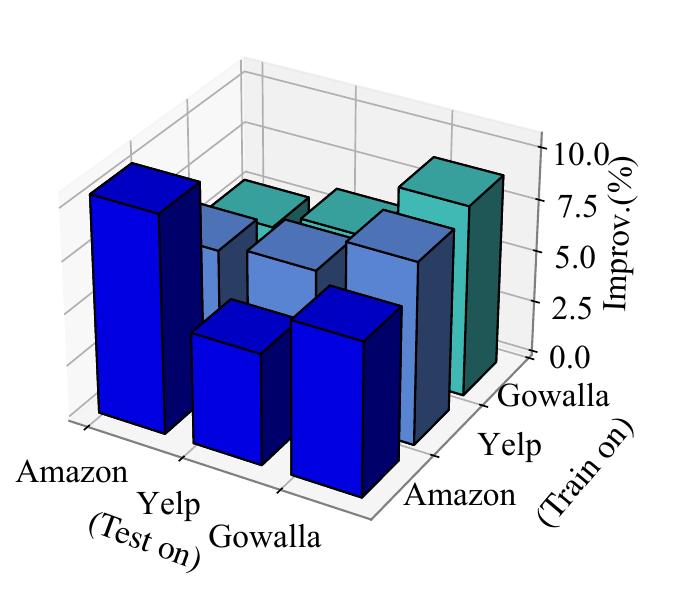}
    \caption{NDCG@20}
\end{subfigure}

\caption{\method{}'s generalization performance.} 
\label{fig:generalization}
\end{figure}

The polynomial Gram filter is parameterized using a series of $\theta_i^{\gamma}$s to represent the convolution kernel. Previous research that employed straightforward, manually designed graph filters \cite{shen2021powerful} has suggested the presence of shared similarities among the optimal response functions across different datasets. This raises the question of whether such generalization of simple filters still holds true for our proposed polynomial generalized Gram filters.

To investigate this, we first train \method{} on three datasets to obtain the corresponding sets of $\theta$ values that characterize the optimal graph kernel. Subsequently, we apply these obtained sets of polynomial filters to other datasets to assess \method{}'s generalization capability. We report the relative improvement of the transferred results over the performance of \method{} with randomly initialized parameters. From results in Figure \ref{fig:generalization} we can observe that, compared to randomly setting the convolution kernel, the parameters obtained from other datasets remain effective to a certain extent.

Additionally, it's noteworthy that the filter transferred from another dataset does not achieve comparable performance compared to the filter trained on the current dataset. This suggests that there are distinctive features within interaction signals specific to different datasets. This observation also underscores the limitations of manually crafted filters, as they may struggle to approximate distinct optimal filter structures tailored to individual datasets.

\subsubsection{Visualization of Polynomial Filter}

\begin{figure}
\centering
\begin{subfigure}{0.49\linewidth}
    \includegraphics[width=\linewidth]{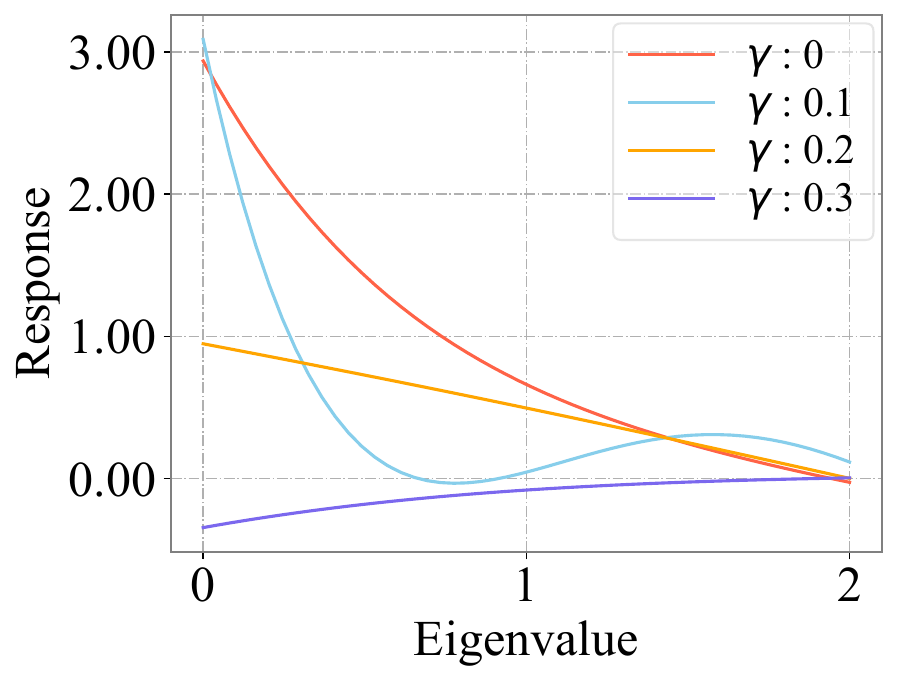}
    \caption{Yelp2018}
\end{subfigure}
\begin{subfigure}{0.49\linewidth}
    \includegraphics[width=\linewidth]{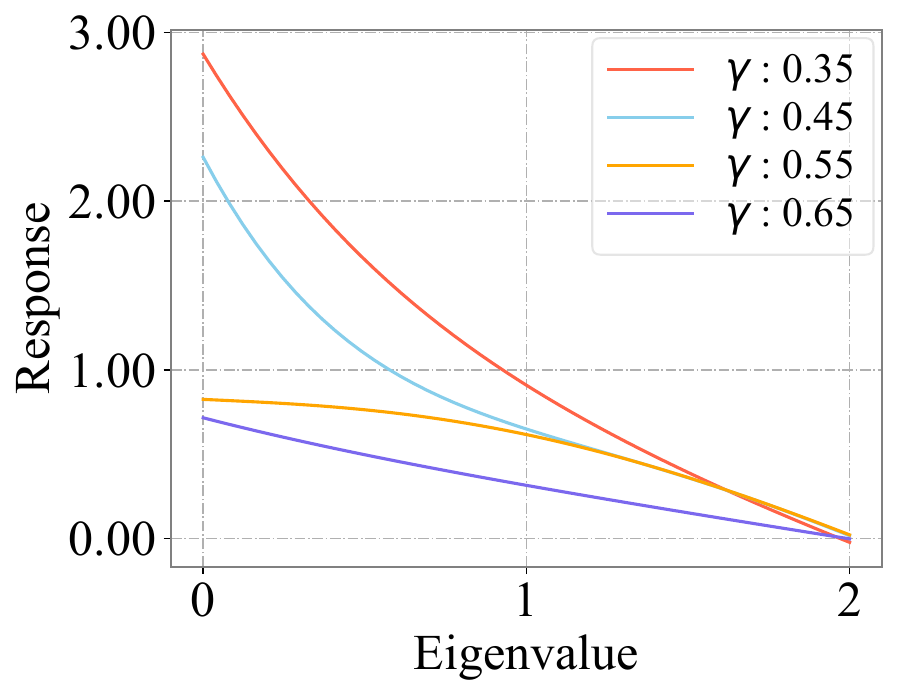}
    \caption{Gowalla}
\end{subfigure}

\caption{Visualization of \method{}'s convolution kernel.} 
\label{fig:visualization}
\end{figure}

The filtering process of the polynomial Gram filter depicts a response function that captures characteristic spectral features of input interaction signals. To intuitively show the learned response function, we visualize the response function and the corresponding polynomials based on the parameters obtained from the Gowalla and Yelp2018 datasets. 

From the visualization results in Figure \ref{fig:visualization} we can observe that the polynomial filter exhibits a more complex structure in its response function. This complexity arises from the distinct eigenspace structures introduced by the multiple generalized normalizations of the Gram matrix, enabling the polynomial convolution kernels to collaborate with each other and form a more expressive convolution structure.
In general, the convolution kernel tends to resemble a low-pass filter, which aligns with empirical observations from prior research. However, there can still be negative gain function corresponding to specific $\gamma$s, which complements the general response function. More specifically, convolution kernels with smaller $\gamma$ values tend to produce a steeper response function.

\section{Conclusion}

In this work, we dive into the expressiveness upper-bound of previous node embedding-based CF methods. To break the limits of existing methods, we propose \method{}, which is a graph filter-based model that solves the CF problem via graph signal processing. Specifically, \method{} incorporates a polynomial graph filter, wherein the convolution kernel is constructed from a sequence of generalized normalization of Gram matrices. This design allows it to effectively capture distinctive spectral characteristics originating from diverse eigenspace structures. A graph optimization-based objective function as well as a Bayesian ranking objective function are jointly used to optimize \method{}, as an approximation of the optimal graph filter that can recover the missing interactions. An extensive set of experiments demonstrates that the proposed \method{} attains state-of-the-art performance across three widely used CF datasets. Furthermore, our additional investigations verify the functionality of each component within \method{}.

% \begin{acks}
% \end{acks}

\bibliographystyle{ACM-Reference-Format}
\bibliography{main}


\begin{thebibliography}{48}

%%% the Web address.

\ifx \showCODEN    \undefined \def \showCODEN     #1{\unskip}     \fi
\ifx \showDOI      \undefined \def \showDOI       #1{#1}\fi
\ifx \showISBNx    \undefined \def \showISBNx     #1{\unskip}     \fi
\ifx \showISBNxiii \undefined \def \showISBNxiii  #1{\unskip}     \fi
\ifx \showISSN     \undefined \def \showISSN      #1{\unskip}     \fi
\ifx \showLCCN     \undefined \def \showLCCN      #1{\unskip}     \fi
\ifx \shownote     \undefined \def \shownote      #1{#1}          \fi
\ifx \showarticletitle \undefined \def \showarticletitle #1{#1}   \fi
\ifx \showURL      \undefined \def \showURL       {\relax}        \fi
% invisible to TeX
\providecommand\bibfield[2]{#2}
\providecommand\bibinfo[2]{#2}
\providecommand\natexlab[1]{#1}
\providecommand\showeprint[2][]{arXiv:#2}

\bibitem[Brin(1998)]%
        {brin1998pagerank}
\bibfield{author}{\bibinfo{person}{Sergey Brin}.} \bibinfo{year}{1998}\natexlab{}.
\newblock \showarticletitle{The PageRank citation ranking: bringing order to the web}.
\newblock \bibinfo{journal}{\emph{Proceedings of ASIS, 1998}}  \bibinfo{volume}{98} (\bibinfo{year}{1998}), \bibinfo{pages}{161--172}.
\newblock


\bibitem[Brin and Page(1998)]%
        {brin1998anatomy}
\bibfield{author}{\bibinfo{person}{Sergey Brin} {and} \bibinfo{person}{Lawrence Page}.} \bibinfo{year}{1998}\natexlab{}.
\newblock \showarticletitle{The anatomy of a large-scale hypertextual web search engine}.
\newblock \bibinfo{journal}{\emph{Computer networks and ISDN systems}} \bibinfo{volume}{30}, \bibinfo{number}{1-7} (\bibinfo{year}{1998}), \bibinfo{pages}{107--117}.
\newblock


\bibitem[Cai et~al\mbox{.}(2023)]%
        {cai2023lightgcl}
\bibfield{author}{\bibinfo{person}{Xuheng Cai}, \bibinfo{person}{Chao Huang}, \bibinfo{person}{Lianghao Xia}, {and} \bibinfo{person}{Xubin Ren}.} \bibinfo{year}{2023}\natexlab{}.
\newblock \showarticletitle{LightGCL: Simple Yet Effective Graph Contrastive Learning for Recommendation}.
\newblock \bibinfo{journal}{\emph{arXiv preprint arXiv:2302.08191}} (\bibinfo{year}{2023}).
\newblock


\bibitem[Chen et~al\mbox{.}(2023)]%
        {chen2023graph}
\bibfield{author}{\bibinfo{person}{Chao Chen}, \bibinfo{person}{Haoyu Geng}, \bibinfo{person}{Gang Zeng}, \bibinfo{person}{Zhaobing Han}, \bibinfo{person}{Hua Chai}, \bibinfo{person}{Xiaokang Yang}, {and} \bibinfo{person}{Junchi Yan}.} \bibinfo{year}{2023}\natexlab{}.
\newblock \showarticletitle{Graph Signal Sampling for Inductive One-Bit Matrix Completion: a Closed-form Solution}.
\newblock \bibinfo{journal}{\emph{arXiv preprint arXiv:2302.03933}} (\bibinfo{year}{2023}).
\newblock


\bibitem[Chen et~al\mbox{.}(2020)]%
        {chen2020efficient}
\bibfield{author}{\bibinfo{person}{Chong Chen}, \bibinfo{person}{Min Zhang}, \bibinfo{person}{Yongfeng Zhang}, \bibinfo{person}{Yiqun Liu}, {and} \bibinfo{person}{Shaoping Ma}.} \bibinfo{year}{2020}\natexlab{}.
\newblock \showarticletitle{Efficient neural matrix factorization without sampling for recommendation}.
\newblock \bibinfo{journal}{\emph{ACM Transactions on Information Systems (TOIS)}} \bibinfo{volume}{38}, \bibinfo{number}{2} (\bibinfo{year}{2020}), \bibinfo{pages}{1--28}.
\newblock


\bibitem[Covington et~al\mbox{.}(2016)]%
        {covington2016deep}
\bibfield{author}{\bibinfo{person}{Paul Covington}, \bibinfo{person}{Jay Adams}, {and} \bibinfo{person}{Emre Sargin}.} \bibinfo{year}{2016}\natexlab{}.
\newblock \showarticletitle{Deep neural networks for youtube recommendations}. In \bibinfo{booktitle}{\emph{Proceedings of the 10th ACM conference on recommender systems}}. \bibinfo{pages}{191--198}.
\newblock


\bibitem[Defferrard et~al\mbox{.}(2016)]%
        {defferrard2016convolutional}
\bibfield{author}{\bibinfo{person}{Micha{\"e}l Defferrard}, \bibinfo{person}{Xavier Bresson}, {and} \bibinfo{person}{Pierre Vandergheynst}.} \bibinfo{year}{2016}\natexlab{}.
\newblock \showarticletitle{Convolutional neural networks on graphs with fast localized spectral filtering}.
\newblock \bibinfo{journal}{\emph{Advances in neural information processing systems}}  \bibinfo{volume}{29} (\bibinfo{year}{2016}).
\newblock


\bibitem[Dong et~al\mbox{.}(2020)]%
        {dong2020graph}
\bibfield{author}{\bibinfo{person}{Xiaowen Dong}, \bibinfo{person}{Dorina Thanou}, \bibinfo{person}{Laura Toni}, \bibinfo{person}{Michael Bronstein}, {and} \bibinfo{person}{Pascal Frossard}.} \bibinfo{year}{2020}\natexlab{}.
\newblock \showarticletitle{Graph signal processing for machine learning: A review and new perspectives}.
\newblock \bibinfo{journal}{\emph{IEEE Signal processing magazine}} \bibinfo{volume}{37}, \bibinfo{number}{6} (\bibinfo{year}{2020}), \bibinfo{pages}{117--127}.
\newblock


\bibitem[Ebesu et~al\mbox{.}(2018)]%
        {ebesu2018collaborative}
\bibfield{author}{\bibinfo{person}{Travis Ebesu}, \bibinfo{person}{Bin Shen}, {and} \bibinfo{person}{Yi Fang}.} \bibinfo{year}{2018}\natexlab{}.
\newblock \showarticletitle{Collaborative memory network for recommendation systems}. In \bibinfo{booktitle}{\emph{The 41st international ACM SIGIR conference on research \& development in information retrieval}}. \bibinfo{pages}{515--524}.
\newblock


\bibitem[Fan et~al\mbox{.}(2019a)]%
        {fan2019graph}
\bibfield{author}{\bibinfo{person}{Wenqi Fan}, \bibinfo{person}{Yao Ma}, \bibinfo{person}{Qing Li}, \bibinfo{person}{Yuan He}, \bibinfo{person}{Eric Zhao}, \bibinfo{person}{Jiliang Tang}, {and} \bibinfo{person}{Dawei Yin}.} \bibinfo{year}{2019}\natexlab{a}.
\newblock \showarticletitle{Graph neural networks for social recommendation}. In \bibinfo{booktitle}{\emph{The world wide web conference}}. \bibinfo{pages}{417--426}.
\newblock


\bibitem[Fan et~al\mbox{.}(2019b)]%
        {fan2019deep}
\bibfield{author}{\bibinfo{person}{Wenqi Fan}, \bibinfo{person}{Yao Ma}, \bibinfo{person}{Dawei Yin}, \bibinfo{person}{Jianping Wang}, \bibinfo{person}{Jiliang Tang}, {and} \bibinfo{person}{Qing Li}.} \bibinfo{year}{2019}\natexlab{b}.
\newblock \showarticletitle{Deep social collaborative filtering}. In \bibinfo{booktitle}{\emph{Proceedings of the 13th ACM conference on recommender systems}}. \bibinfo{pages}{305--313}.
\newblock


\bibitem[Fu et~al\mbox{.}(2022)]%
        {fu2022revisiting}
\bibfield{author}{\bibinfo{person}{Hao-Ming Fu}, \bibinfo{person}{Patrick Poirson}, \bibinfo{person}{Kwot~Sin Lee}, {and} \bibinfo{person}{Chen Wang}.} \bibinfo{year}{2022}\natexlab{}.
\newblock \showarticletitle{Revisiting Neighborhood-based Link Prediction for Collaborative Filtering}. In \bibinfo{booktitle}{\emph{Companion Proceedings of the Web Conference 2022}}. \bibinfo{pages}{1009--1018}.
\newblock


\bibitem[Gasteiger et~al\mbox{.}(2018)]%
        {gasteiger2018predict}
\bibfield{author}{\bibinfo{person}{Johannes Gasteiger}, \bibinfo{person}{Aleksandar Bojchevski}, {and} \bibinfo{person}{Stephan G{\"u}nnemann}.} \bibinfo{year}{2018}\natexlab{}.
\newblock \showarticletitle{Predict then propagate: Graph neural networks meet personalized pagerank}.
\newblock \bibinfo{journal}{\emph{arXiv preprint arXiv:1810.05997}} (\bibinfo{year}{2018}).
\newblock


\bibitem[Goldberg et~al\mbox{.}(1992)]%
        {goldberg1992using}
\bibfield{author}{\bibinfo{person}{David Goldberg}, \bibinfo{person}{David Nichols}, \bibinfo{person}{Brian~M Oki}, {and} \bibinfo{person}{Douglas Terry}.} \bibinfo{year}{1992}\natexlab{}.
\newblock \showarticletitle{Using collaborative filtering to weave an information tapestry}.
\newblock \bibinfo{journal}{\emph{Commun. ACM}} \bibinfo{volume}{35}, \bibinfo{number}{12} (\bibinfo{year}{1992}), \bibinfo{pages}{61--70}.
\newblock


\bibitem[Gopalan et~al\mbox{.}(2015)]%
        {gopalan2015scalable}
\bibfield{author}{\bibinfo{person}{Prem Gopalan}, \bibinfo{person}{Jake~M Hofman}, {and} \bibinfo{person}{David~M Blei}.} \bibinfo{year}{2015}\natexlab{}.
\newblock \showarticletitle{Scalable Recommendation with Hierarchical Poisson Factorization.}. In \bibinfo{booktitle}{\emph{UAI}}. \bibinfo{pages}{326--335}.
\newblock


\bibitem[Gori et~al\mbox{.}(2007)]%
        {gori2007itemrank}
\bibfield{author}{\bibinfo{person}{Marco Gori}, \bibinfo{person}{Augusto Pucci}, \bibinfo{person}{Via Roma}, {and} \bibinfo{person}{I Siena}.} \bibinfo{year}{2007}\natexlab{}.
\newblock \showarticletitle{Itemrank: A random-walk based scoring algorithm for recommender engines.}. In \bibinfo{booktitle}{\emph{IJCAI}}, Vol.~\bibinfo{volume}{7}. \bibinfo{pages}{2766--2771}.
\newblock


\bibitem[Guo et~al\mbox{.}(2023)]%
        {guo2023manipulating}
\bibfield{author}{\bibinfo{person}{Jiayan Guo}, \bibinfo{person}{Lun Du}, \bibinfo{person}{Xu Chen}, \bibinfo{person}{Xiaojun Ma}, \bibinfo{person}{Qiang Fu}, \bibinfo{person}{Shi Han}, \bibinfo{person}{Dongmei Zhang}, {and} \bibinfo{person}{Yan Zhang}.} \bibinfo{year}{2023}\natexlab{}.
\newblock \showarticletitle{On Manipulating Signals of User-Item Graph: A Jacobi Polynomial-based Graph Collaborative Filtering}.
\newblock \bibinfo{journal}{\emph{arXiv preprint arXiv:2306.03624}} (\bibinfo{year}{2023}).
\newblock


\bibitem[Hamilton et~al\mbox{.}(2017)]%
        {hamilton2017inductive}
\bibfield{author}{\bibinfo{person}{Will Hamilton}, \bibinfo{person}{Zhitao Ying}, {and} \bibinfo{person}{Jure Leskovec}.} \bibinfo{year}{2017}\natexlab{}.
\newblock \showarticletitle{Inductive representation learning on large graphs}.
\newblock \bibinfo{journal}{\emph{Advances in neural information processing systems}}  \bibinfo{volume}{30} (\bibinfo{year}{2017}).
\newblock


\bibitem[He et~al\mbox{.}(2021)]%
        {he2021bernnet}
\bibfield{author}{\bibinfo{person}{Mingguo He}, \bibinfo{person}{Zhewei Wei}, \bibinfo{person}{Hongteng Xu}, {et~al\mbox{.}}} \bibinfo{year}{2021}\natexlab{}.
\newblock \showarticletitle{Bernnet: Learning arbitrary graph spectral filters via bernstein approximation}.
\newblock \bibinfo{journal}{\emph{Advances in Neural Information Processing Systems}}  \bibinfo{volume}{34} (\bibinfo{year}{2021}), \bibinfo{pages}{14239--14251}.
\newblock


\bibitem[He et~al\mbox{.}(2020)]%
        {he2020lightgcn}
\bibfield{author}{\bibinfo{person}{Xiangnan He}, \bibinfo{person}{Kuan Deng}, \bibinfo{person}{Xiang Wang}, \bibinfo{person}{Yan Li}, \bibinfo{person}{Yongdong Zhang}, {and} \bibinfo{person}{Meng Wang}.} \bibinfo{year}{2020}\natexlab{}.
\newblock \showarticletitle{Lightgcn: Simplifying and powering graph convolution network for recommendation}. In \bibinfo{booktitle}{\emph{Proceedings of the 43rd International ACM SIGIR conference on research and development in Information Retrieval}}. \bibinfo{pages}{639--648}.
\newblock


\bibitem[He et~al\mbox{.}(2016)]%
        {he2016birank}
\bibfield{author}{\bibinfo{person}{Xiangnan He}, \bibinfo{person}{Ming Gao}, \bibinfo{person}{Min-Yen Kan}, {and} \bibinfo{person}{Dingxian Wang}.} \bibinfo{year}{2016}\natexlab{}.
\newblock \showarticletitle{Birank: Towards ranking on bipartite graphs}.
\newblock \bibinfo{journal}{\emph{IEEE Transactions on Knowledge and Data Engineering}} \bibinfo{volume}{29}, \bibinfo{number}{1} (\bibinfo{year}{2016}), \bibinfo{pages}{57--71}.
\newblock


\bibitem[He et~al\mbox{.}(2017)]%
        {he2017neural}
\bibfield{author}{\bibinfo{person}{Xiangnan He}, \bibinfo{person}{Lizi Liao}, \bibinfo{person}{Hanwang Zhang}, \bibinfo{person}{Liqiang Nie}, \bibinfo{person}{Xia Hu}, {and} \bibinfo{person}{Tat-Seng Chua}.} \bibinfo{year}{2017}\natexlab{}.
\newblock \showarticletitle{Neural collaborative filtering}. In \bibinfo{booktitle}{\emph{Proceedings of the 26th international conference on world wide web}}. \bibinfo{pages}{173--182}.
\newblock


\bibitem[Ju et~al\mbox{.}(2023)]%
        {ju2023comprehensive}
\bibfield{author}{\bibinfo{person}{Wei Ju}, \bibinfo{person}{Zheng Fang}, \bibinfo{person}{Yiyang Gu}, \bibinfo{person}{Zequn Liu}, \bibinfo{person}{Qingqing Long}, \bibinfo{person}{Ziyue Qiao}, \bibinfo{person}{Yifang Qin}, \bibinfo{person}{Jianhao Shen}, \bibinfo{person}{Fang Sun}, \bibinfo{person}{Zhiping Xiao}, {et~al\mbox{.}}} \bibinfo{year}{2023}\natexlab{}.
\newblock \showarticletitle{A Comprehensive Survey on Deep Graph Representation Learning}.
\newblock \bibinfo{journal}{\emph{arXiv preprint arXiv:2304.05055}} (\bibinfo{year}{2023}).
\newblock


\bibitem[Kipf and Welling(2016)]%
        {kipf2016semi}
\bibfield{author}{\bibinfo{person}{Thomas~N Kipf} {and} \bibinfo{person}{Max Welling}.} \bibinfo{year}{2016}\natexlab{}.
\newblock \showarticletitle{Semi-supervised classification with graph convolutional networks}.
\newblock \bibinfo{journal}{\emph{arXiv preprint arXiv:1609.02907}} (\bibinfo{year}{2016}).
\newblock


\bibitem[Koren et~al\mbox{.}(2009)]%
        {koren2009matrix}
\bibfield{author}{\bibinfo{person}{Yehuda Koren}, \bibinfo{person}{Robert Bell}, {and} \bibinfo{person}{Chris Volinsky}.} \bibinfo{year}{2009}\natexlab{}.
\newblock \showarticletitle{Matrix factorization techniques for recommender systems}.
\newblock \bibinfo{journal}{\emph{Computer}} \bibinfo{volume}{42}, \bibinfo{number}{8} (\bibinfo{year}{2009}), \bibinfo{pages}{30--37}.
\newblock


\bibitem[Liu et~al\mbox{.}(2023)]%
        {liu2023personalized}
\bibfield{author}{\bibinfo{person}{Jiahao Liu}, \bibinfo{person}{Dongsheng Li}, \bibinfo{person}{Hansu Gu}, \bibinfo{person}{Tun Lu}, \bibinfo{person}{Peng Zhang}, \bibinfo{person}{Li Shang}, {and} \bibinfo{person}{Ning Gu}.} \bibinfo{year}{2023}\natexlab{}.
\newblock \showarticletitle{Personalized Graph Signal Processing for Collaborative Filtering}. In \bibinfo{booktitle}{\emph{Proceedings of the ACM Web Conference 2023}}. \bibinfo{pages}{1264--1272}.
\newblock


\bibitem[Liu et~al\mbox{.}(2022)]%
        {liu2022revisiting}
\bibfield{author}{\bibinfo{person}{Nian Liu}, \bibinfo{person}{Xiao Wang}, \bibinfo{person}{Deyu Bo}, \bibinfo{person}{Chuan Shi}, {and} \bibinfo{person}{Jian Pei}.} \bibinfo{year}{2022}\natexlab{}.
\newblock \showarticletitle{Revisiting graph contrastive learning from the perspective of graph spectrum}.
\newblock \bibinfo{journal}{\emph{Advances in Neural Information Processing Systems}}  \bibinfo{volume}{35} (\bibinfo{year}{2022}), \bibinfo{pages}{2972--2983}.
\newblock


\bibitem[Mao et~al\mbox{.}(2021a)]%
        {mao2021simplex}
\bibfield{author}{\bibinfo{person}{Kelong Mao}, \bibinfo{person}{Jieming Zhu}, \bibinfo{person}{Jinpeng Wang}, \bibinfo{person}{Quanyu Dai}, \bibinfo{person}{Zhenhua Dong}, \bibinfo{person}{Xi Xiao}, {and} \bibinfo{person}{Xiuqiang He}.} \bibinfo{year}{2021}\natexlab{a}.
\newblock \showarticletitle{SimpleX: A simple and strong baseline for collaborative filtering}. In \bibinfo{booktitle}{\emph{Proceedings of the 30th ACM International Conference on Information \& Knowledge Management}}. \bibinfo{pages}{1243--1252}.
\newblock


\bibitem[Mao et~al\mbox{.}(2021b)]%
        {mao2021ultragcn}
\bibfield{author}{\bibinfo{person}{Kelong Mao}, \bibinfo{person}{Jieming Zhu}, \bibinfo{person}{Xi Xiao}, \bibinfo{person}{Biao Lu}, \bibinfo{person}{Zhaowei Wang}, {and} \bibinfo{person}{Xiuqiang He}.} \bibinfo{year}{2021}\natexlab{b}.
\newblock \showarticletitle{UltraGCN: ultra simplification of graph convolutional networks for recommendation}. In \bibinfo{booktitle}{\emph{Proceedings of the 30th ACM International Conference on Information \& Knowledge Management}}. \bibinfo{pages}{1253--1262}.
\newblock


\bibitem[Mnih and Salakhutdinov(2007)]%
        {mnih2007probabilistic}
\bibfield{author}{\bibinfo{person}{Andriy Mnih} {and} \bibinfo{person}{Russ~R Salakhutdinov}.} \bibinfo{year}{2007}\natexlab{}.
\newblock \showarticletitle{Probabilistic matrix factorization}.
\newblock \bibinfo{journal}{\emph{Advances in neural information processing systems}}  \bibinfo{volume}{20} (\bibinfo{year}{2007}).
\newblock


\bibitem[Qin et~al\mbox{.}(2023a)]%
        {qin2023learning}
\bibfield{author}{\bibinfo{person}{Yifang Qin}, \bibinfo{person}{Wei Ju}, \bibinfo{person}{Hongjun Wu}, \bibinfo{person}{Xiao Luo}, {and} \bibinfo{person}{Ming Zhang}.} \bibinfo{year}{2023}\natexlab{a}.
\newblock \showarticletitle{Learning Graph ODE for Continuous-Time Sequential Recommendation}.
\newblock \bibinfo{journal}{\emph{arXiv preprint arXiv:2304.07042}} (\bibinfo{year}{2023}).
\newblock


\bibitem[Qin et~al\mbox{.}(2023b)]%
        {qin2023disenpoi}
\bibfield{author}{\bibinfo{person}{Yifang Qin}, \bibinfo{person}{Yifan Wang}, \bibinfo{person}{Fang Sun}, \bibinfo{person}{Wei Ju}, \bibinfo{person}{Xuyang Hou}, \bibinfo{person}{Zhe Wang}, \bibinfo{person}{Jia Cheng}, \bibinfo{person}{Jun Lei}, {and} \bibinfo{person}{Ming Zhang}.} \bibinfo{year}{2023}\natexlab{b}.
\newblock \showarticletitle{DisenPOI: Disentangling Sequential and Geographical Influence for Point-of-Interest Recommendation}. In \bibinfo{booktitle}{\emph{Proceedings of the Sixteenth ACM International Conference on Web Search and Data Mining}}. \bibinfo{pages}{508--516}.
\newblock


\bibitem[Qiu et~al\mbox{.}(2020)]%
        {qiu2020gag}
\bibfield{author}{\bibinfo{person}{Ruihong Qiu}, \bibinfo{person}{Hongzhi Yin}, \bibinfo{person}{Zi Huang}, {and} \bibinfo{person}{Tong Chen}.} \bibinfo{year}{2020}\natexlab{}.
\newblock \showarticletitle{Gag: Global attributed graph neural network for streaming session-based recommendation}. In \bibinfo{booktitle}{\emph{Proceedings of the 43rd International ACM SIGIR Conference on Research and Development in Information Retrieval}}. \bibinfo{pages}{669--678}.
\newblock


\bibitem[Ramakrishna et~al\mbox{.}(2020)]%
        {ramakrishna2020user}
\bibfield{author}{\bibinfo{person}{Raksha Ramakrishna}, \bibinfo{person}{Hoi-To Wai}, {and} \bibinfo{person}{Anna Scaglione}.} \bibinfo{year}{2020}\natexlab{}.
\newblock \showarticletitle{A user guide to low-pass graph signal processing and its applications: Tools and applications}.
\newblock \bibinfo{journal}{\emph{IEEE Signal Processing Magazine}} \bibinfo{volume}{37}, \bibinfo{number}{6} (\bibinfo{year}{2020}), \bibinfo{pages}{74--85}.
\newblock


\bibitem[Rendle et~al\mbox{.}(2012)]%
        {rendle2012bpr}
\bibfield{author}{\bibinfo{person}{Steffen Rendle}, \bibinfo{person}{Christoph Freudenthaler}, \bibinfo{person}{Zeno Gantner}, {and} \bibinfo{person}{Lars Schmidt-Thieme}.} \bibinfo{year}{2012}\natexlab{}.
\newblock \showarticletitle{BPR: Bayesian personalized ranking from implicit feedback}.
\newblock \bibinfo{journal}{\emph{arXiv preprint arXiv:1205.2618}} (\bibinfo{year}{2012}).
\newblock


\bibitem[Shen et~al\mbox{.}(2021)]%
        {shen2021powerful}
\bibfield{author}{\bibinfo{person}{Yifei Shen}, \bibinfo{person}{Yongji Wu}, \bibinfo{person}{Yao Zhang}, \bibinfo{person}{Caihua Shan}, \bibinfo{person}{Jun Zhang}, \bibinfo{person}{B~Khaled Letaief}, {and} \bibinfo{person}{Dongsheng Li}.} \bibinfo{year}{2021}\natexlab{}.
\newblock \showarticletitle{How powerful is graph convolution for recommendation?}. In \bibinfo{booktitle}{\emph{Proceedings of the 30th ACM international conference on information \& knowledge management}}. \bibinfo{pages}{1619--1629}.
\newblock


\bibitem[Shuman et~al\mbox{.}(2013)]%
        {shuman2013emerging}
\bibfield{author}{\bibinfo{person}{David~I Shuman}, \bibinfo{person}{Sunil~K Narang}, \bibinfo{person}{Pascal Frossard}, \bibinfo{person}{Antonio Ortega}, {and} \bibinfo{person}{Pierre Vandergheynst}.} \bibinfo{year}{2013}\natexlab{}.
\newblock \showarticletitle{The emerging field of signal processing on graphs: Extending high-dimensional data analysis to networks and other irregular domains}.
\newblock \bibinfo{journal}{\emph{IEEE signal processing magazine}} \bibinfo{volume}{30}, \bibinfo{number}{3} (\bibinfo{year}{2013}), \bibinfo{pages}{83--98}.
\newblock


\bibitem[Steck(2019)]%
        {steck2019embarrassingly}
\bibfield{author}{\bibinfo{person}{Harald Steck}.} \bibinfo{year}{2019}\natexlab{}.
\newblock \showarticletitle{Embarrassingly shallow autoencoders for sparse data}. In \bibinfo{booktitle}{\emph{The World Wide Web Conference}}. \bibinfo{pages}{3251--3257}.
\newblock


\bibitem[Tay et~al\mbox{.}(2018)]%
        {tay2018latent}
\bibfield{author}{\bibinfo{person}{Yi Tay}, \bibinfo{person}{Luu Anh~Tuan}, {and} \bibinfo{person}{Siu~Cheung Hui}.} \bibinfo{year}{2018}\natexlab{}.
\newblock \showarticletitle{Latent relational metric learning via memory-based attention for collaborative ranking}. In \bibinfo{booktitle}{\emph{Proceedings of the 2018 world wide web conference}}. \bibinfo{pages}{729--739}.
\newblock


\bibitem[Wang et~al\mbox{.}(2018)]%
        {wang2018billion}
\bibfield{author}{\bibinfo{person}{Jizhe Wang}, \bibinfo{person}{Pipei Huang}, \bibinfo{person}{Huan Zhao}, \bibinfo{person}{Zhibo Zhang}, \bibinfo{person}{Binqiang Zhao}, {and} \bibinfo{person}{Dik~Lun Lee}.} \bibinfo{year}{2018}\natexlab{}.
\newblock \showarticletitle{Billion-scale commodity embedding for e-commerce recommendation in alibaba}. In \bibinfo{booktitle}{\emph{Proceedings of the 24th ACM SIGKDD international conference on knowledge discovery \& data mining}}. \bibinfo{pages}{839--848}.
\newblock


\bibitem[Wang et~al\mbox{.}(2019)]%
        {wang2019neural}
\bibfield{author}{\bibinfo{person}{Xiang Wang}, \bibinfo{person}{Xiangnan He}, \bibinfo{person}{Meng Wang}, \bibinfo{person}{Fuli Feng}, {and} \bibinfo{person}{Tat-Seng Chua}.} \bibinfo{year}{2019}\natexlab{}.
\newblock \showarticletitle{Neural graph collaborative filtering}. In \bibinfo{booktitle}{\emph{Proceedings of the 42nd international ACM SIGIR conference on Research and development in Information Retrieval}}. \bibinfo{pages}{165--174}.
\newblock


\bibitem[Wang et~al\mbox{.}(2020)]%
        {wang2020disentangled}
\bibfield{author}{\bibinfo{person}{Xiang Wang}, \bibinfo{person}{Hongye Jin}, \bibinfo{person}{An Zhang}, \bibinfo{person}{Xiangnan He}, \bibinfo{person}{Tong Xu}, {and} \bibinfo{person}{Tat-Seng Chua}.} \bibinfo{year}{2020}\natexlab{}.
\newblock \showarticletitle{Disentangled graph collaborative filtering}. In \bibinfo{booktitle}{\emph{Proceedings of the 43rd international ACM SIGIR conference on research and development in information retrieval}}. \bibinfo{pages}{1001--1010}.
\newblock


\bibitem[Wang and Zhang(2022)]%
        {wang2022powerful}
\bibfield{author}{\bibinfo{person}{Xiyuan Wang} {and} \bibinfo{person}{Muhan Zhang}.} \bibinfo{year}{2022}\natexlab{}.
\newblock \showarticletitle{How powerful are spectral graph neural networks}. In \bibinfo{booktitle}{\emph{International Conference on Machine Learning}}. PMLR, \bibinfo{pages}{23341--23362}.
\newblock


\bibitem[Wang et~al\mbox{.}(2023)]%
        {wang2023collaboration}
\bibfield{author}{\bibinfo{person}{Yu Wang}, \bibinfo{person}{Yuying Zhao}, \bibinfo{person}{Yi Zhang}, {and} \bibinfo{person}{Tyler Derr}.} \bibinfo{year}{2023}\natexlab{}.
\newblock \showarticletitle{Collaboration-Aware Graph Convolutional Network for Recommender Systems}. In \bibinfo{booktitle}{\emph{Proceedings of the ACM Web Conference 2023}}. \bibinfo{pages}{91--101}.
\newblock


\bibitem[Wang et~al\mbox{.}(2022)]%
        {wang2022graph}
\bibfield{author}{\bibinfo{person}{Zhaobo Wang}, \bibinfo{person}{Yanmin Zhu}, \bibinfo{person}{Qiaomei Zhang}, \bibinfo{person}{Haobing Liu}, \bibinfo{person}{Chunyang Wang}, {and} \bibinfo{person}{Tong Liu}.} \bibinfo{year}{2022}\natexlab{}.
\newblock \showarticletitle{Graph-enhanced spatial-temporal network for next POI recommendation}.
\newblock \bibinfo{journal}{\emph{ACM Transactions on Knowledge Discovery from Data (TKDD)}} \bibinfo{volume}{16}, \bibinfo{number}{6} (\bibinfo{year}{2022}), \bibinfo{pages}{1--21}.
\newblock


\bibitem[Wu et~al\mbox{.}(2021)]%
        {wu2021self}
\bibfield{author}{\bibinfo{person}{Jiancan Wu}, \bibinfo{person}{Xiang Wang}, \bibinfo{person}{Fuli Feng}, \bibinfo{person}{Xiangnan He}, \bibinfo{person}{Liang Chen}, \bibinfo{person}{Jianxun Lian}, {and} \bibinfo{person}{Xing Xie}.} \bibinfo{year}{2021}\natexlab{}.
\newblock \showarticletitle{Self-supervised graph learning for recommendation}. In \bibinfo{booktitle}{\emph{Proceedings of the 44th international ACM SIGIR conference on research and development in information retrieval}}. \bibinfo{pages}{726--735}.
\newblock


\bibitem[Wu et~al\mbox{.}(2019)]%
        {wu2019session}
\bibfield{author}{\bibinfo{person}{Shu Wu}, \bibinfo{person}{Yuyuan Tang}, \bibinfo{person}{Yanqiao Zhu}, \bibinfo{person}{Liang Wang}, \bibinfo{person}{Xing Xie}, {and} \bibinfo{person}{Tieniu Tan}.} \bibinfo{year}{2019}\natexlab{}.
\newblock \showarticletitle{Session-based recommendation with graph neural networks}. In \bibinfo{booktitle}{\emph{Proceedings of the AAAI conference on artificial intelligence}}, Vol.~\bibinfo{volume}{33}. \bibinfo{pages}{346--353}.
\newblock


\bibitem[Xia et~al\mbox{.}(2022)]%
        {xia2022fire}
\bibfield{author}{\bibinfo{person}{Jiafeng Xia}, \bibinfo{person}{Dongsheng Li}, \bibinfo{person}{Hansu Gu}, \bibinfo{person}{Jiahao Liu}, \bibinfo{person}{Tun Lu}, {and} \bibinfo{person}{Ning Gu}.} \bibinfo{year}{2022}\natexlab{}.
\newblock \showarticletitle{FIRE: Fast incremental recommendation with graph signal processing}. In \bibinfo{booktitle}{\emph{Proceedings of the ACM Web Conference 2022}}. \bibinfo{pages}{2360--2369}.
\newblock


\end{thebibliography}
% \input{Content/7-appendix}

% \appendix

\end{document}